\documentclass[nofootinbib,aps,11pt]{revtex4-1}
\usepackage{graphicx}
\usepackage{hyperref}
\usepackage{cancel}
\usepackage{amssymb}
\usepackage{textcomp}
\usepackage{amsmath}
\usepackage{bm}
\usepackage{times}
\usepackage{epsfig}
\usepackage{color}
\usepackage{float}

\begin{document}
\title{\LARGE Sterile Neutrino Portal Dark Matter with $Z_3$ Symmetry}
\bigskip
\author{An Liu$^1$}
\author{Zhi-Long Han$^1$}
\email{sps\_hanzl@ujn.edu.cn}
\author{Yi Jin$^{1,2}$}
\author{Honglei Li$^1$}
\affiliation{
$^1$School of Physics and Technology, University of Jinan, Jinan, Shandong 250022, China
\\
$^2$Guangxi Key Laboratory of Nuclear Physics and Nuclear Technology, Guangxi Normal University, Guilin, Guangxi 541004, China}
\date{\today}

\begin{abstract}
In this paper, we consider the sterile neutrino portal dark matter with $Z_3$ symmetry. This model further extends the canonical type-I seesaw  with a fermion singlet $\chi$ and a scalar singlet $\phi$. Under the $Z_3$ symmetry, the dark sector transforms as $\chi\to e^{i2\pi/3}\chi, \phi\to e^{i2\pi/3}\phi$, while the standard model particles and the sterile neutrino $N$ transform trivially. Besides the interactions as $y_{N} \phi \bar{\chi}N$ and $\lambda_{H\phi} (H^\dag H) (\phi^\dag \phi)$ allowed in the $Z_2$ symmetry, the $Z_3$ symmetry also introduces two new terms, i.e., $y_\chi \phi \overline{\chi^{c}} \chi$ and $\mu\phi^3/2$. These new interactions induce additional semi-annihilation processes as $\chi\chi\to N\chi$ and $\phi\phi\to h\phi$ for WIMP dark matter. We then perform a comprehensive analysis of the phenomenology of this $Z_3$ symmetric model. Viable parameter space is explored under the constraints from dark matter relic density, Higgs invisible decay, indirect and direct detection for both fermion and scalar dark matter. We find that the semi-annihilation channels $\chi\chi\to N\chi$ and $\phi\phi\to N\chi$ can lead to quite different phenomena from the $Z_2$ symmetric model, which provides a viable pathway to distinguish these two kinds of model. 
\end{abstract}

\maketitle

\section{Introduction}
The identity of particle dark matter (DM)  and the explanation for the tiny mass of neutrinos remain outstanding questions in particle physics, garnering attention as crucial topics in current research. Their common origin presents the possibility of future exploration into new physics beyond the standard model. The Weakly Interacting Massive Particle (WIMP) is the most promising dark matter candidate \cite{Arcadi:2017kky}. However, this scenario, such as the extensively studied Higgs portal \cite{McDonald:1993ex,Burgess:2000yq,Cline:2013gha} and $Z'$ portal model \cite{Alves:2013tqa,Alves:2015pea,Rodejohann:2015lca,Klasen:2016qux}, usually suffer stringent constraints from direct detection. Therefore, a new interaction portal for the WIMP dark sector should be considered.

Sterile neutrinos $N$ are introduced to generate the tiny neutrino mass via the canonical Type-I seesaw mechanism \cite{Minkowski:1977sc,Mohapatra:1979ia}. For a proper mixing angle with active neutrinos, the keV scale sterile neutrino can serve as a decaying dark matter \cite{Dodelson:1993je,Boyarsky:2018tvu}. Then the radiative decay $N\to \nu \gamma$ leads to an observable signature at X-ray telescopes \cite{Roach:2019ctw}, which is able to explain the tentative 3.5 keV line signal \cite{Boyarsky:2014ska}. However, the parameter space for sterile neutrino dark matter now is tightly constrained \cite{Roach:2022lgo}.  If the sterile neutrinos $N$ are charged under the dark group, the lightest sterile neutrino becomes a stable dark matter. In this scenario, the tree-level Type-I seesaw is also forbidden by the dark group, then light neutrino mass could be generated via the radiative mechanism \cite{Ma:2006km,Ma:2007gq,Ding:2016wbd,Liu:2022byu}.

On the other hand, the electroweak scale sterile neutrino $N$ is an ideal messenger between the dark sector and the standard model \cite{Ballett:2019pyw,Blennow:2019fhy,Falkowski:2009yz,Falkowski:2011xh,GonzalezMacias:2015rxl,Gonzalez-Macias:2016vxy,Liu:2020mxj,Coito:2022kif,Liu:2022rst}. This is facilitated through the new Yukawa coupling $y_N \phi \bar{\chi}N$, which enables the secluded channel $\phi\phi/\chi\chi\to NN$, providing an additional annihilation pathway for the WIMP dark matter. In particular, this scenario features a relatively small nucleon scattering cross section, and permits the indirect detection of observable gamma-ray signals \cite{Pospelov:2007mp,Folgado:2018qlv,Tang:2015coo,Campos:2017odj,Batell:2017rol,Batell:2017cmf}. For an electroweak scale dark matter annihilating via the sterile neutrino portal, it is still allowed by direct detection and is hopefully probed by indirect detection in the near future. Meanwhile, the sterile neutrino portal dark matter produced through the freeze-in mechanism is also extensively studied \cite{Falkowski:2017uya,Becker:2018rve,Chianese:2018dsz,Bandyopadhyay:2020qpn,Cheng:2020gut,Chang:2021ose,Barman:2022scg,Liu:2022cct}.

The interactions between the dark sector and the standard model particles are typically governed by the dark group, such as the well-studied $Z_2$ \cite{Escudero:2016ksa} or $U(1)_{B-L}$ symmetry \cite{Escudero:2016tzx}. Although the simplest $Z_2$ symmetry  has demonstrated success in dark matter with a simplified phenomenology, more sophisticated dark groups, e.g. $Z_N(N\geq3)$ \cite{Batell:2010bp}, $A_4$ \cite{Hirsch:2010ru} and $SU(2)$  symmetry \cite{Diaz-Cruz:2010czr}, are also options. For instance, an alternative explanation for the observed relic abundance of dark matter through semi-annihilation may be achieved by introducing the next simplest $Z_3$ symmetry, leading to a lower bound on the direct detection cross section \cite{Belanger:2012vp,Belanger:2012zr}.

In this paper, we consider the sterile neutrino portal dark matter with $Z_3$ symmetry \cite{Bandyopadhyay:2022tsf,Ghosh:2023ocl}. This model includes a fermion singlet $\chi$ and a scalar singlet $\phi$, both of which transform non-trivially under the exact $Z_3$ symmetry as $\chi\to e^{i2\pi/3}\chi, \phi\to e^{i2\pi/3}\phi$. The standard model particles and the sterile neutrinos are not charged under the imposed $Z_3$ symmetry. Compared with the $Z_2$ symmetry, the $Z_3$ symmetry allows two new interaction terms, i.e., $y_\chi \phi \overline{\chi^{c}} \chi$ and $\mu\phi^3/2$, which would lead to new annihilation channels of dark matter. The semi-annihilation of fermion dark matter via the process $\chi\chi\to N\chi $ in the framework of effective field theory has been considered in Ref.~\cite{Bandyopadhyay:2022tsf}. Focusing on the self-interaction of dark scalar $\phi$, Ref.~\cite{Ghosh:2023ocl} studies the non-thermal production of dark matter $\chi$ by the late time decay $\phi\to \chi \nu$. In this paper, we perform a comprehensive analysis of WIMP dark matter for both scalar and fermion scenarios. The non-thermal production of dark matter  will be considered in a separate paper \cite{Liu:2023hzl}.

This paper is structured into several sections. In Sec. \ref{SEC:TM}, we introduce the sterile neutrino portal dark matter model with $Z_3$ symmetry. In Sec. \ref{SEC:RD}, we illustrate the evolution of the dark matter abundance under certain scenarios. Then we perform a random scan to obtain the viable parameter space for correct relic abundance. In Sec. \ref{SEC:HID}, we calculate the branching ratio of Higgs invisible decay. In Sec. \ref{SEC:ID}, we explore the indirect detection constraints of dark matter. In Sec. \ref{SEC:DD}, we compute the direct detection cross section of dark matter. Finally, in Sec. \ref{SEC:CL}, we provide concluding remarks for our study.

\section{The Model}\label{SEC:TM}

This model further extends the Type-I seesaw with a dark sector under the $Z_3$ symmetry. Sterile neutrinos $N$ are introduced to generate tiny neutrino mass. In this paper, we consider the electroweak scale $N$ in order to accommodate WIMP dark matter. The dark sector consists of a scalar singlet $\phi$ and a fermion singlet $\chi$. Under the dark $Z_3$ symmetry, the dark sector transforms as $\chi\to e^{i2\pi/3}\chi, \phi\to e^{i2\pi/3}\phi$, while the standard model particles and the sterile neutrinos transform trivially. The lightest particle in the dark sector serves as dark matter. In this paper, both fermion and scalar dark matter will be considered.

The Yukawa interaction takes the form of
\begin{equation}\label{Eq:Yuk}
	-\mathcal{L}_Y=\left(y_\nu \overline{L} \tilde{H} N + y_{N} \phi \bar{\chi}N +  h.c.\right) + y_\chi \phi \overline{\chi^c}\chi ,
\end{equation}
where $L$ is the left-handed lepton doublet and $H$ is the Higgs doublet with $\tilde{H}=i\sigma_2 H^*$. Light neutrino mass is generated by the Type-I seesaw as
\begin{equation}
	m_\nu=-\frac{v^2}{2}y_\nu m_N^{-1} y_\nu^T,
\end{equation}
where $v=246$ GeV is the vacuum expectation value of the Higgs field. For electroweak scale sterile neutrinos, the mixing angle with light neutrino $\theta$ is at the order of $\sqrt{m_\nu/m_N}\sim10^{-6}$, which is far below current collider limits \cite{Abdullahi:2022jlv}.

The scalar potential under the exact $Z_3$ symmetry is
\begin{equation}\label{la1}
	V=-\mu_H^2 H^\dag H + \mu_\phi^2 \phi^\dag \phi + \lambda_H (H^\dag H)^2 + \lambda_\phi (\phi^\dag \phi)^2 + \lambda_{H\phi} (H^\dag H) (\phi^\dag \phi) + \left(\frac{\mu}{2}\phi^3+h.c.\right),
\end{equation}
where all the parameters are taken to be real. After the electroweak symmetry breaking, the physical mass of dark scalar $\phi$ is $m_\phi^2=\mu^2_\phi+ \lambda_{H \phi} v^2/2$. The scalar potential in Equation \eqref{la1} must have a finite minimum to prevent unbounded energy, which requires \cite{Belanger:2012zr}
\begin{equation}
	\lambda_{H} > 0, \quad \lambda_{\phi} > 0, \quad \lambda_{H\phi} + 2 \sqrt{\lambda_{H} \lambda_{\phi}} > 0.
\end{equation}
Meanwhile, the stability of electroweak vacuum sets an upper bound on the cubic coupling $\mu$ as
\begin{equation}
	\mu\leq 2\sqrt{\lambda_\phi} m_\phi,
\end{equation}
in the limit of small $\lambda_{H \phi}$. In order to maintain the validity of perturbation theory,  $\left|\lambda_{\phi}\right| \leqslant \pi$ and $\left|\lambda_{H\phi}\right| \leqslant 4 \pi$ should be further satisfied. In the following studies, we assume $\mu<3 m_\phi$, which is also allowed by the unitarity constraints \cite{Hektor:2019ote}.

\section{Relic density}\label{SEC:RD}

\begin{figure}[htbp]
	\begin{center}
		\includegraphics[width=0.7\linewidth]{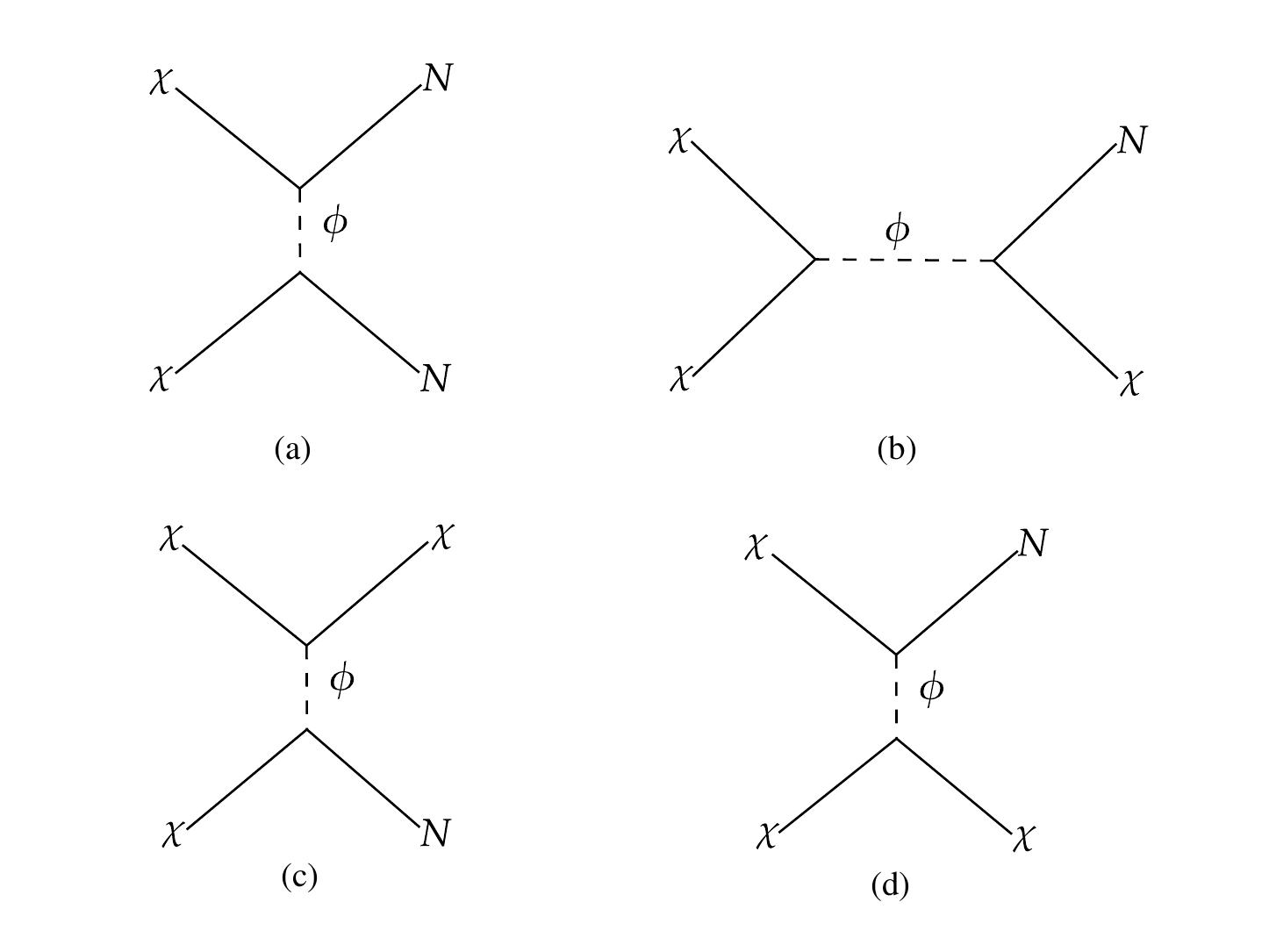}
	\end{center}
	\caption{The dominant annihilation channels of fermion dark matter. Panel (a) is for the secluded channel $\chi\chi\to NN$, while panels (b)-(d) are for the semi-annihilation channel $\chi\chi\to N\chi$. }
	\label{ff}
\end{figure}

\begin{figure}
	\begin{center}
	\includegraphics[width=1\linewidth]{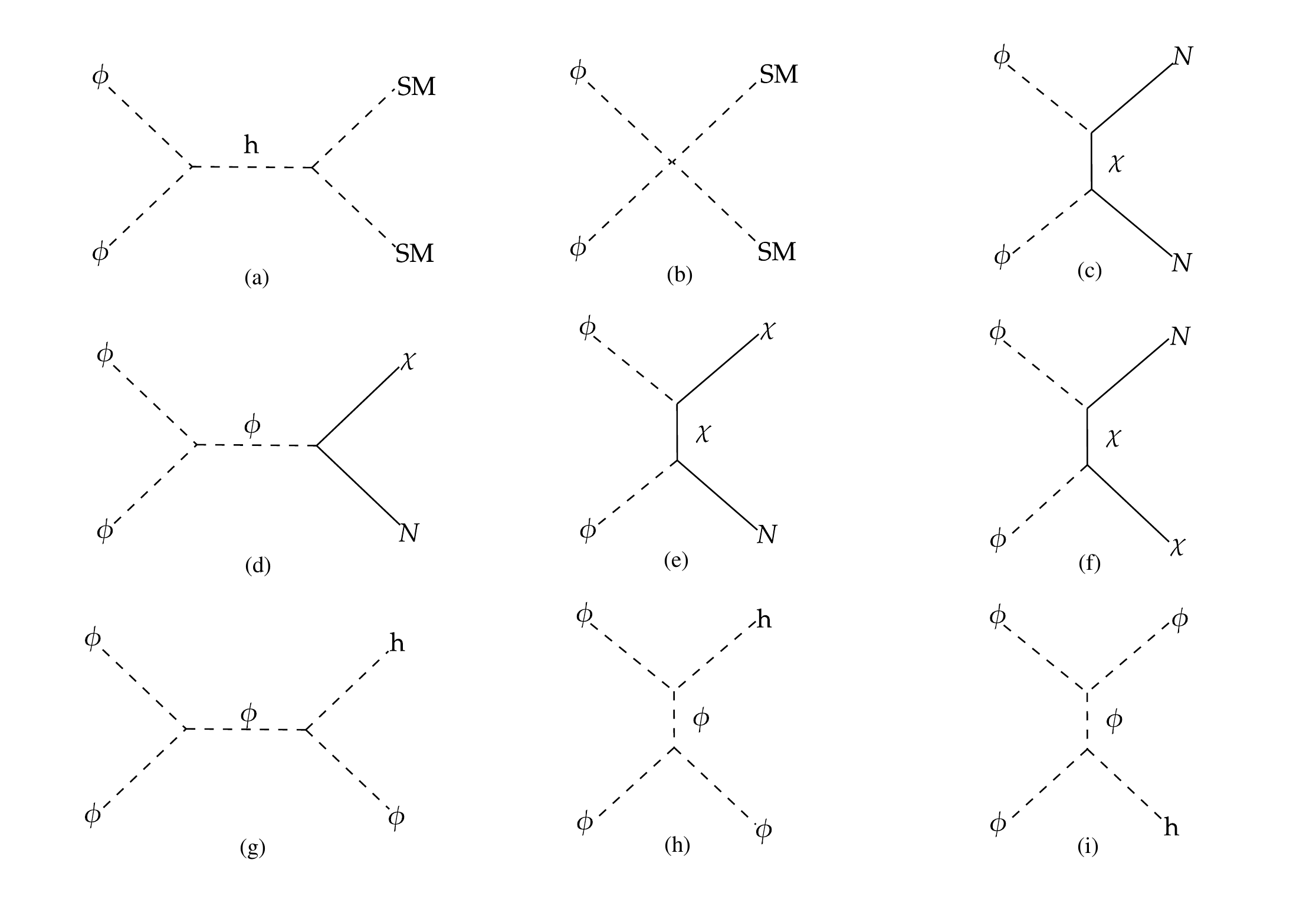}
	\end{center}
	\caption{The dominant annihilation channels of scalar dark matter. Panel (a) and (b) denote the annihilation channels to SM final states.  Panel (c) is for the secluded channel $\phi\phi\to NN$. Panels (d)-(f) are for the fermion semi-annihilation channel $\phi\phi\to N\chi$. Panels (g)-(i) are for the scalar semi-annihilation channel $\phi\phi\to h\phi$. }
	\label{sf}
\end{figure}

In this section, we first discuss the annihilation channels of dark matter. As shown in Figure~\ref{ff} , there are two dominant annihilation channels for the fermion dark matter $\chi$. One is the secluded channel $\chi\chi\to NN$, which also exists in the $Z_2$ symmetry model. The other one is the semi-annihilation channel $\chi\chi\to N\chi$ \cite{DEramo:2010keq}, which is induced by the new Yukawa coupling $y_\chi \phi \overline{\chi^{c}}\chi$ under the $Z_3$ symmetry. In principle, there is also the scalar semi-annihilation channel $\chi\chi\to \phi h$, however the annihilation cross section is $p$-wave suppressed for the Majorana-like Yukawa coupling $y_\chi \phi \overline{\chi^{c}}\chi$ \cite{Cai:2015zza,Guo:2021rre}. For the scalar dark matter $\phi$, there are four kinds of annihilation channels as depicted in Figure~\ref{sf} . Apart from the extensively studied Higgs portal channels $\phi\phi\to \text{SM}$, the secluded channel $\phi\phi\to NN$ is also allowed. Meanwhile, the cubic term $\mu \phi^3$ and the Yukawa coupling $y_\chi \phi \overline{\chi^{c}}\chi$  induce two additional semi-annihilation channels $\phi\phi\to N\chi$ and $\phi\phi\to h\phi$ if kinematically allowed. Comparing with the simplest $Z_3$ scalar singlet dark matter \cite{Hektor:2019ote}, the fermion channel $\phi\phi\to N\chi$ is unique in this model. Therefore, the sterile neutrino portal semi-annihilation channels $\chi\chi\to N\chi$ and $\phi\phi\to N\chi$ provide a viable pathway to distinguish from the other models. It is notable that when masses of the dark sector are nearly degenerate, the co-annihilation channels as $\phi\chi\to h\chi/\phi N$ are also possible. For simplicity, we do not consider such co-annihilation channels in this paper.

\begin{figure}
	\begin{center}
		\includegraphics[width=0.45\linewidth]{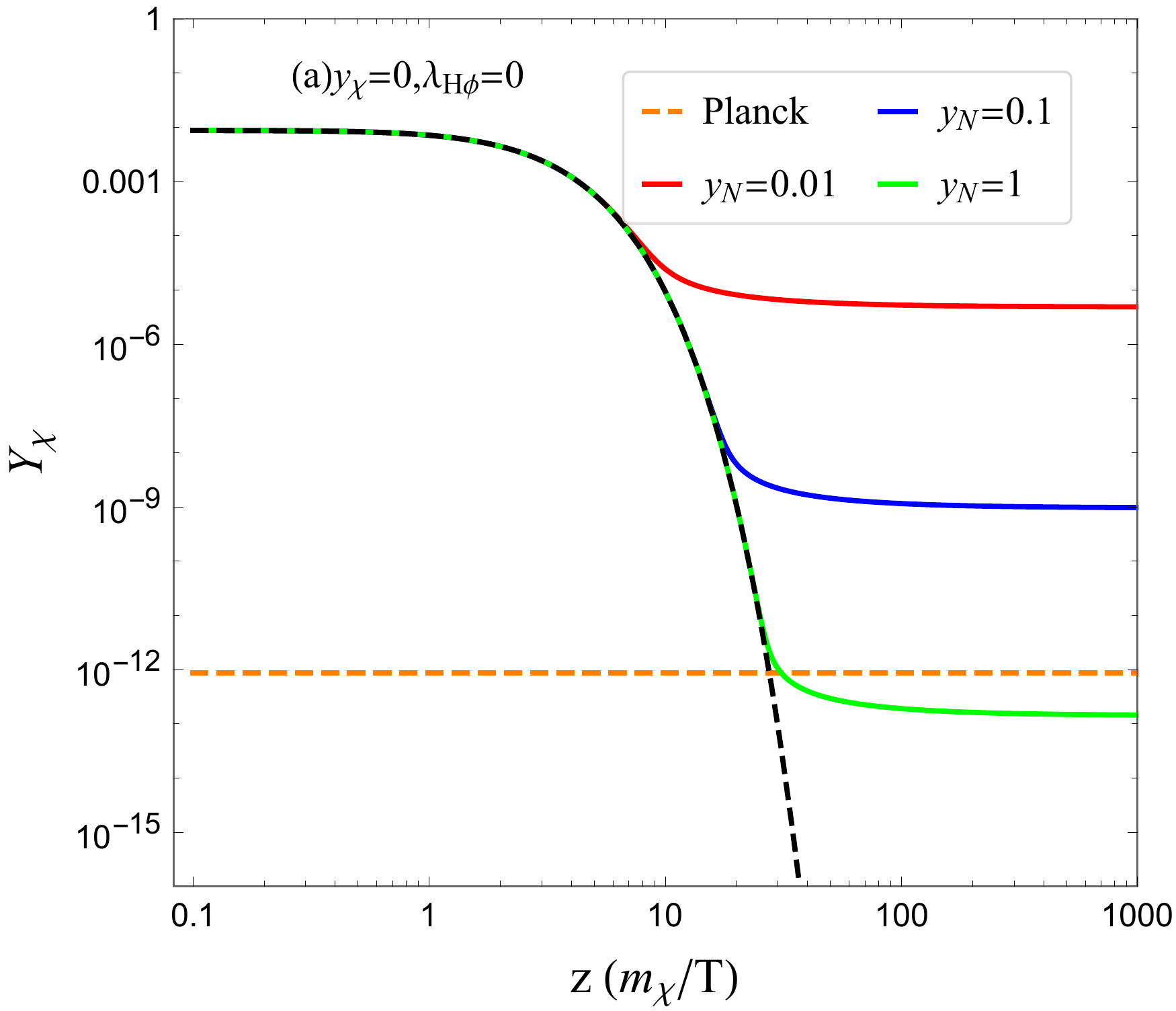}
		\includegraphics[width=0.45\linewidth]{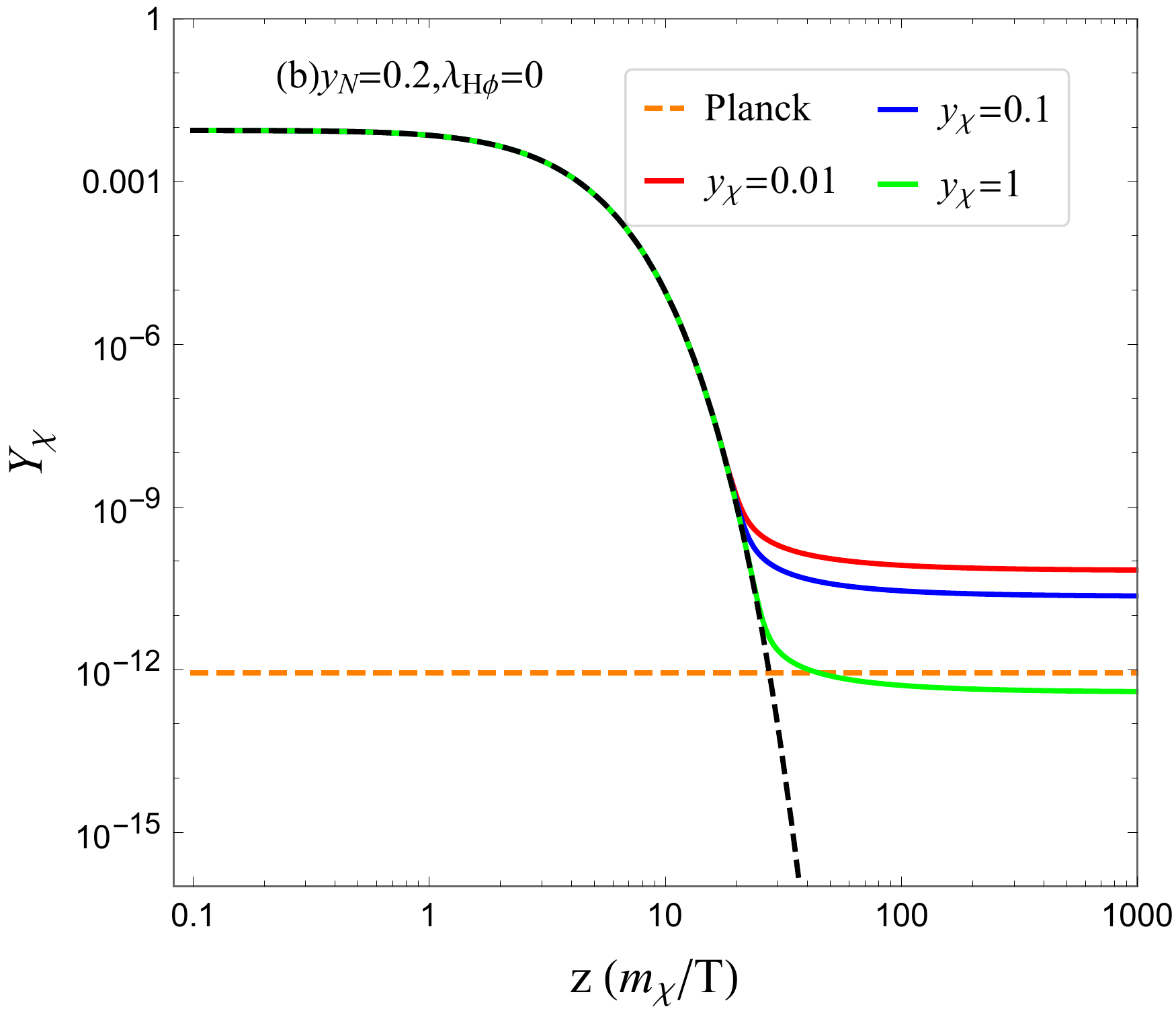}
	\end{center}
	\caption{The evolution of fermion dark matter abundance in different major annihilation channels.  The orange horizontal lines correspond to the Planck observed abundance for $m_{\mathrm{DM}}=500$ GeV. }
	\label{relicF}
\end{figure}

As the WIMP dark matter candidate, it is in thermal equilibrium at the very beginning, and then decouples from the thermal bath when the temperature is low enough. Defining the variable $z=m_\text{DM}/T$, the evolution of fermion dark matter abundance $Y_\chi$ is determined by the Boltzmann equation
\begin{eqnarray}
	\frac{\mathrm{d} Y_\chi}{\mathrm{d} z}=-\frac{\lambda}{z^{2}}\langle\sigma v\rangle_{\chi\chi\to NN}\left(Y_\chi^{2}-(Y_\chi^{\mathrm{eq}})^2\right)-\frac{\lambda}{2z^{2}}\langle\sigma v\rangle_{\chi\chi\to N\chi}\left(Y_\chi^{2}-Y_\chi^{\mathrm{eq}} Y_\chi\right),
\end{eqnarray}
where $\lambda$ is defined as $\lambda \equiv \sqrt{\pi g_{*}/{45}} m_\mathrm{DM} M_{\mathrm{Pl}}$. Here, $g_{*}$ is the effective number of degrees of freedom of the relativistic species and $M_{\mathrm{Pl}}=1.2\times10^{19}$ GeV is the Planck mass. The sterile neutrino $N$ is assumed in thermal equilibrium \cite{Li:2022bpp}. Similarly, the evolution of scalar dark matter is calculated as
\begin{eqnarray}
	\frac{\mathrm{d} Y_\phi}{\mathrm{d} z}&=&-\frac{\lambda}{z^{2}}\langle\sigma v\rangle_{\phi\phi\to \mathrm{SM}}\left(Y_\phi^{2}-(Y_\phi^{\mathrm{eq}})^2\right)
	-\frac{\lambda}{2z^{2}}\langle\sigma v\rangle_{\phi\phi\to h\phi }\left(Y_\phi^{2}-Y_\phi^{\mathrm{eq}} Y_\phi\right)
	\\\nonumber
	&&-\frac{\lambda}{z^{2}}\langle\sigma v\rangle_{\phi\phi\to NN}\left(Y_\phi^{2}-(Y_\phi^{\mathrm{eq}})^2\right)-\frac{\lambda}{2z^{2}}\langle\sigma v\rangle_{\phi\phi\to N\chi}\left(Y_\phi^{2}-\frac{(Y_\phi^{\mathrm{eq}})^2}{Y_\chi^{\mathrm{eq}}} Y_\chi\right)
.
\end{eqnarray}
The thermally averaged cross sections $\langle\sigma v\rangle$ are calculated with MicrOMEGAs \cite{Belanger:2018ccd}.

\begin{figure}
	\begin{center}
		\includegraphics[width=0.45\linewidth,height=0.4\linewidth]{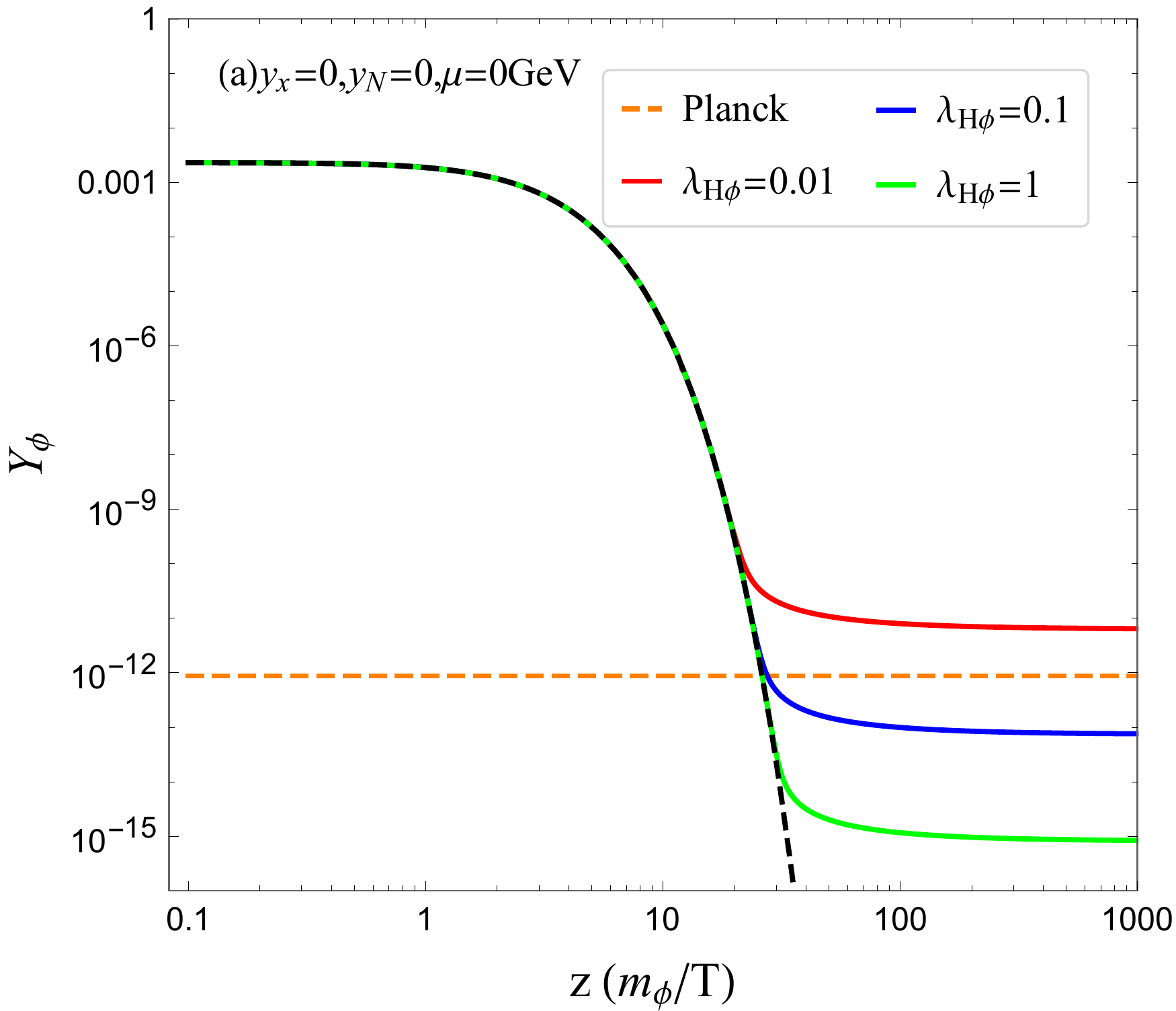}
		\includegraphics[width=0.45\linewidth,height=0.4\linewidth]{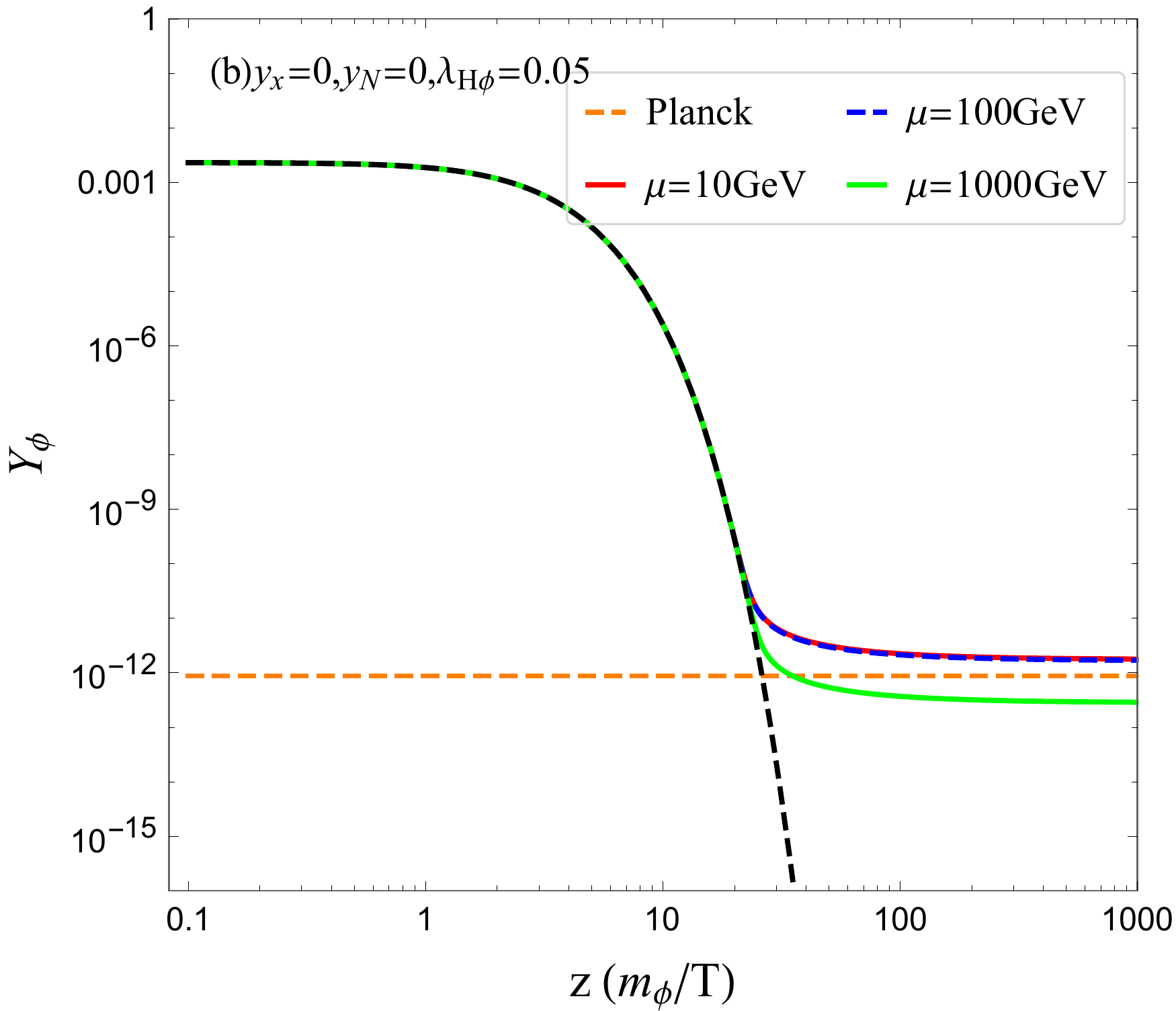}
		\includegraphics[width=0.45\linewidth,height=0.4\linewidth]{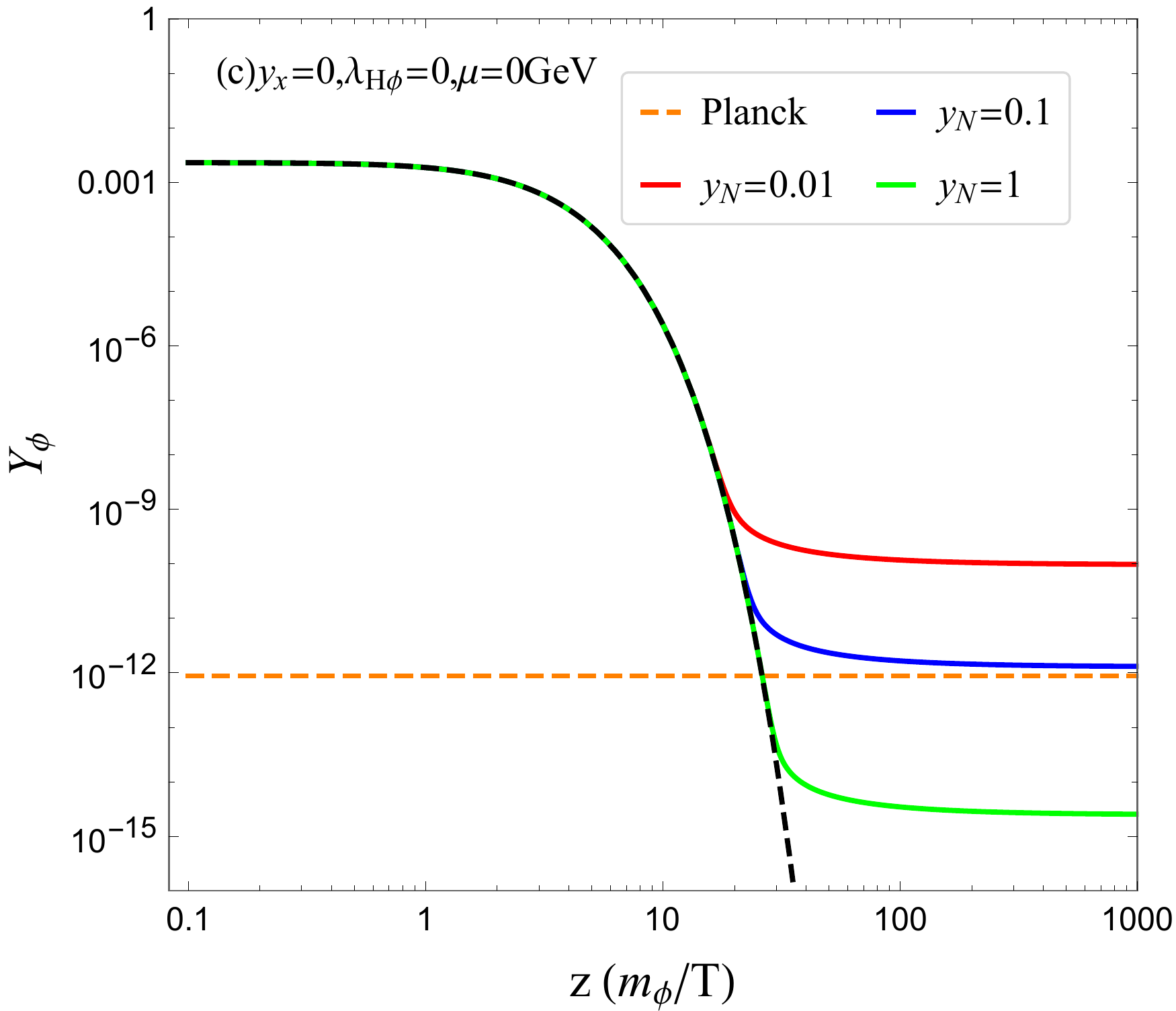}
		\includegraphics[width=0.45\linewidth,height=0.4\linewidth]{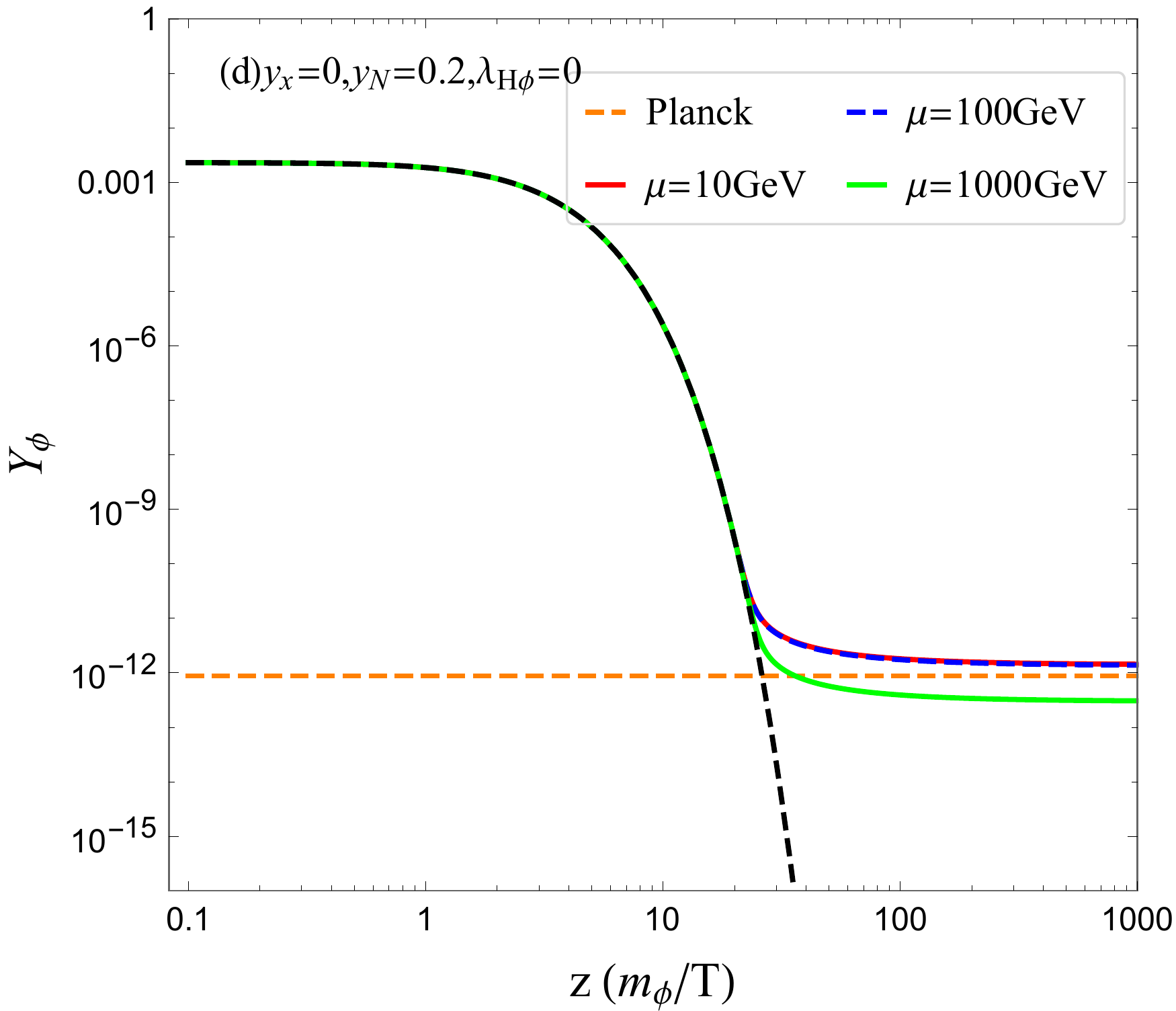}
		\includegraphics[width=0.45\linewidth,height=0.4\linewidth]{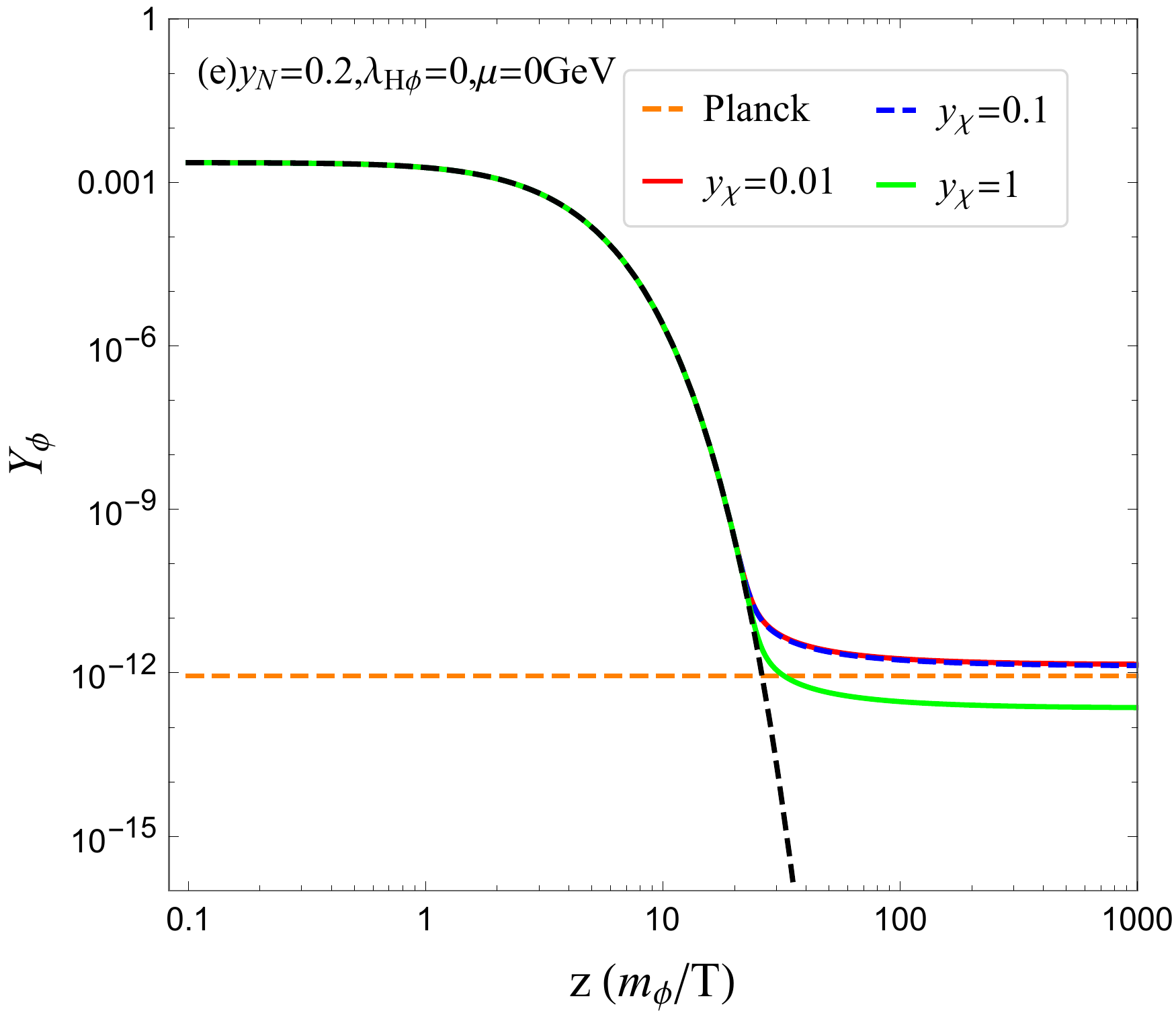}
		\includegraphics[width=0.45\linewidth,height=0.4\linewidth]{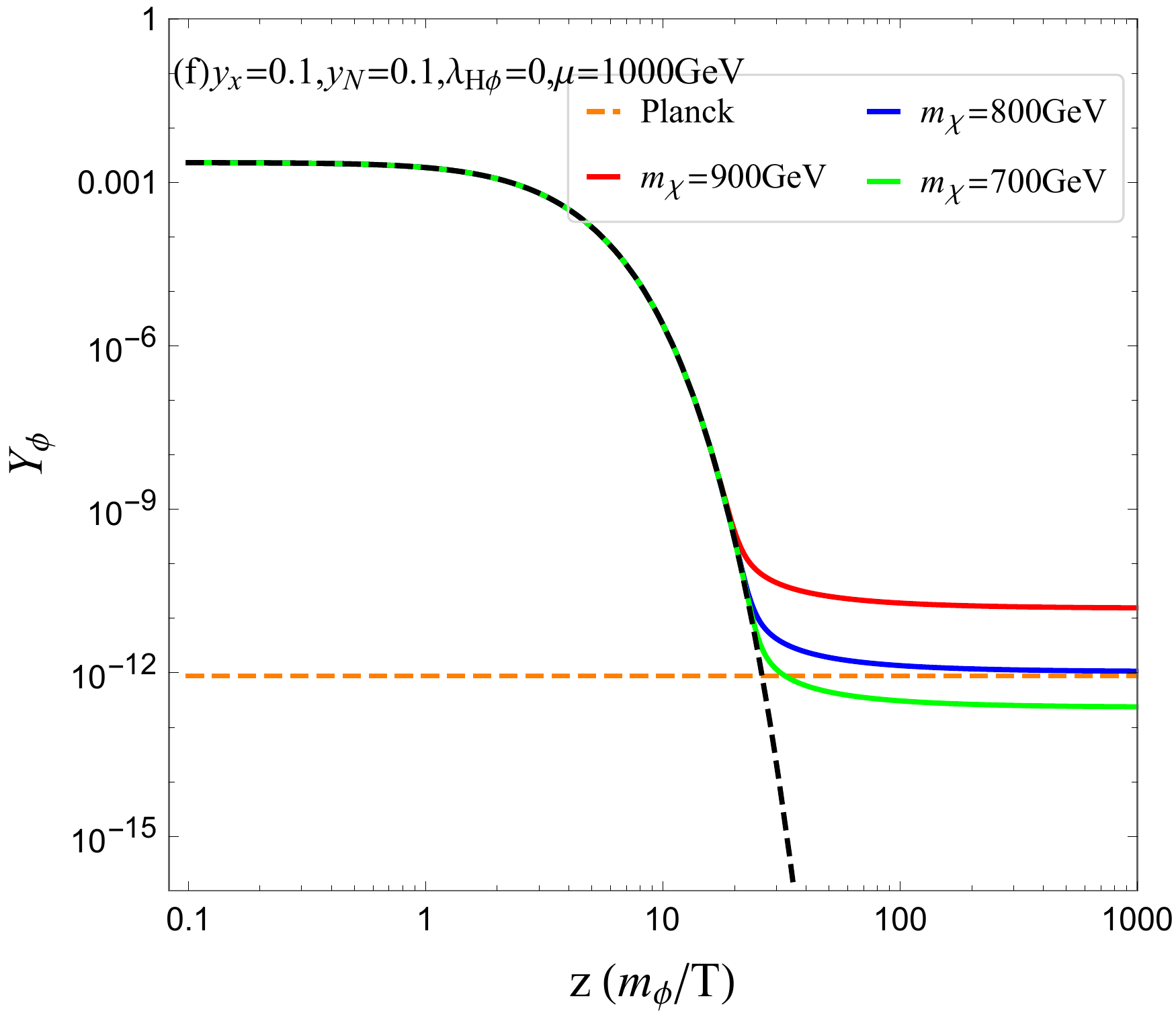}
	\end{center}
	\caption{Same as Figure \ref{relicF}, but for scalar dark matter. }
	\label{relicS}
\end{figure}

Figure~\ref{relicF} and Figure ~\ref{relicS} present the evolution of the dark matter relic abundance via various annihilation channels during the early universe. According to the general definition, the lighter of fermion $\chi$ and scalar $\phi$ is the dark matter candidate. For illustration, we set the dark matter  mass  $m_{\mathrm{DM}}=500$ GeV, the other heavier particle mass $m_{\mathrm{Heavier}}=800$ GeV, and the sterile neutrino mass  $m_N=180$ GeV. The secluded channel $\chi\chi\to NN$ only involves the Yukawa coupling $y_N$, whose impacts on the abundance is shown in panel (a) of Figure~\ref{relicF}. The contribution of semi-annihilation channel $\chi\chi\to N\chi$ is turned off simply by setting $y_\chi=0$. To obtain correct relic abundance, $y_\chi\sim\mathcal{O}(0.1)$ is required when $\chi\chi\to NN$ is the only annihilation channel. For the process $\chi\chi\to N\chi$, both the Yukawa coupling $y_N$ and $y_\chi$ contribute. We then fix $y_N=0.2$, and show the impact of $y_\chi$ in panel (b). The observed relic abundance is reproduced with $y_\chi\sim\mathcal{O}(1)$. Since $\chi\chi\to N N$ is also kinematically allowed, it is clear that when $y_\chi\ll y_N$, the relic abundance is actually determined by the secluded channel.

For the scalar dark matter $\phi$, we first show the impact of $\lambda_{H \phi}$ on the canonical Higgs portal annihilation channels in panel (a) of Figure~\ref{relicS}. Contributions of other kinds of annihilation channels are forbidden by fixing $y_\chi=y_N=0$ and $\mu=0$ GeV. These Higgs portal channels are efficient to obtain the desired abundance with $\lambda_{H \phi}\gtrsim\mathcal{O}(0.01)$. The contribution of scalar semi-annihilation $\phi\phi\to h \phi$ is depicted in panel (b) while setting $\lambda_{H \phi}=0.05$. A relatively large cubic coupling $\mu\gtrsim100$ GeV is required to make this channel the dominant one. In panels (c) to (f), we consider the secluded channel $\phi\phi\to NN$ and fermion semi-annihilation channel $\phi\phi\to N\chi$. Similar to the fermion dark matter scenario, correct abundance is achieved with $y_N\sim\mathcal{O}(0.1)$ or $y_\chi\sim\mathcal{O}(1)$ when $\phi\phi\to NN$ or $\phi\phi\to N\chi$ is the dominant annihilation channel respectively. Different from the fermion dark matter, the $s$-channel of  semi-annihilation $\phi\phi\to N\chi$ is induced by the cubic term $\mu\phi^3$ but not the Yukawa coupling $y_\chi \phi \overline{\chi^{c}}\chi$. The contributions of the $s$- and $t/u$-channels are then separately shown in panels (d) and (e). Since the contribution of $\phi\phi\to N\chi$ is suppressed by the final states phase space in the benchmark points, a larger contribution is possible with lighter $\chi,N$, which is illustrated in panel (f).

Generating the appropriate cosmological relic density, as determined with high accuracy by the Planck experiment $	\Omega_{\rm DM} h^2 = 0.120 \pm 0.001 $ \cite{Planck:2018vyg}, is a crucial prerequisite for a viable dark matter candidate.  With a seesaw related mixing angle $\theta\sim\sqrt{m_\nu/m_N}$, the lifetime of sterile neutrino $\tau_N$ would be longer than $\mathcal{O}(0.1)$ s for $m_N<1$ GeV, which is excluded by Big Bang Nucleosynthesis \cite{Ruchayskiy:2012si}. So we assume $m_N>1$~GeV in this paper, and then perform a random scan to explore the following dark sector parameter space:
\begin{equation}\label{Eq:PR}
	y_{\chi,N}\in[10^{-4},1],\lambda_{H\phi}\in[10^{-6},1], \mu/m_\phi\in[0,3],m_{\chi,\phi}\in[1,10^3]~\text{GeV}.
\end{equation}

\begin{figure}
	\begin{center}
		\includegraphics[width=1\linewidth]{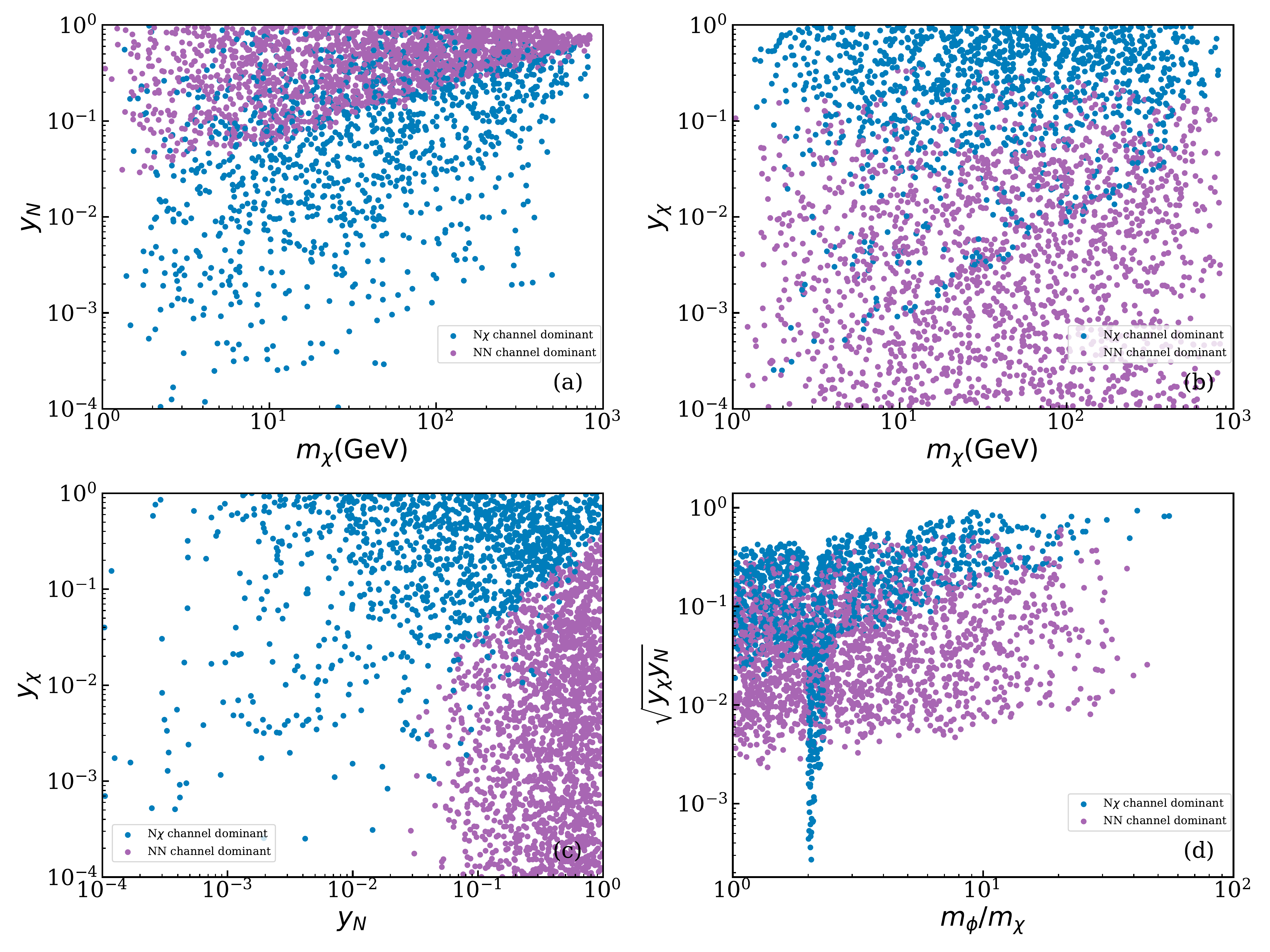}
	\end{center}
	\caption{Distributions of samples with correct relic density for fermion dark matter. The purple and blue points denote that the dominant annihilation channel is $\chi\chi\to NN$ and $\chi\chi\to N\chi$ respectively.}
	\label{Rf}
\end{figure}

\begin{figure}
	\begin{center}
	\includegraphics[width=1\linewidth]{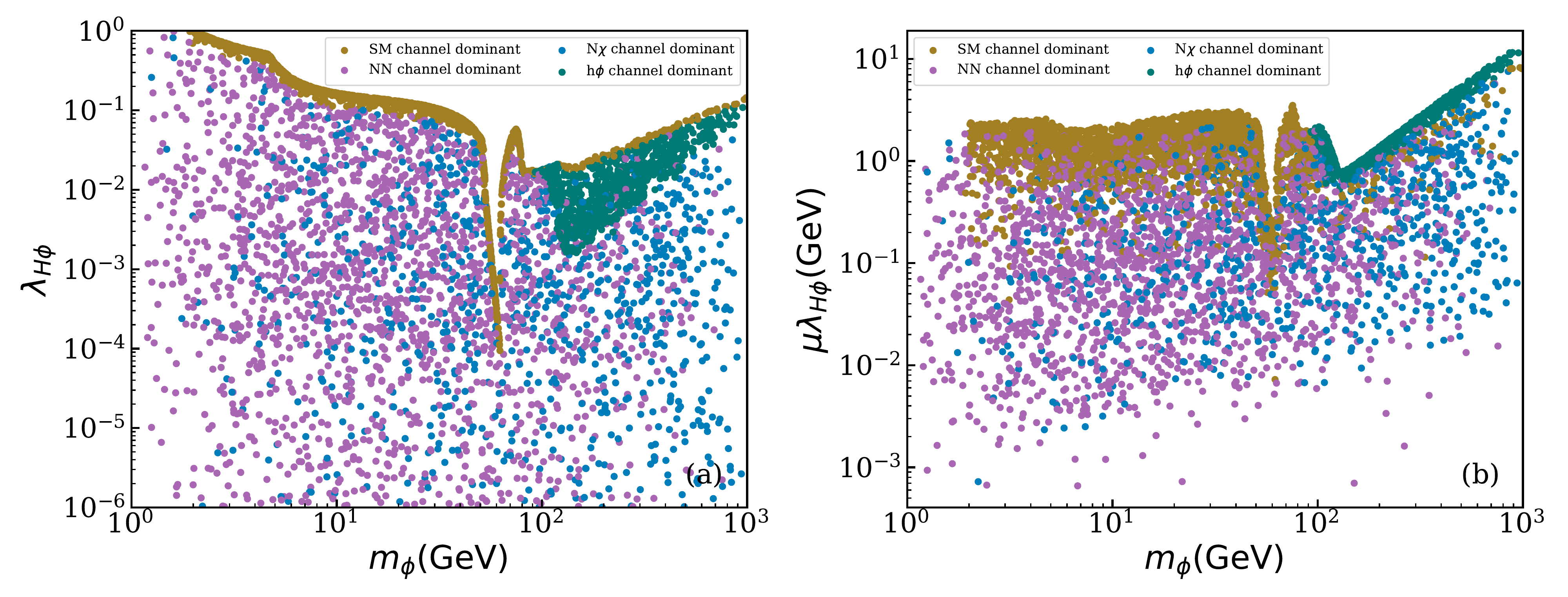}
	\includegraphics[width=1\linewidth]{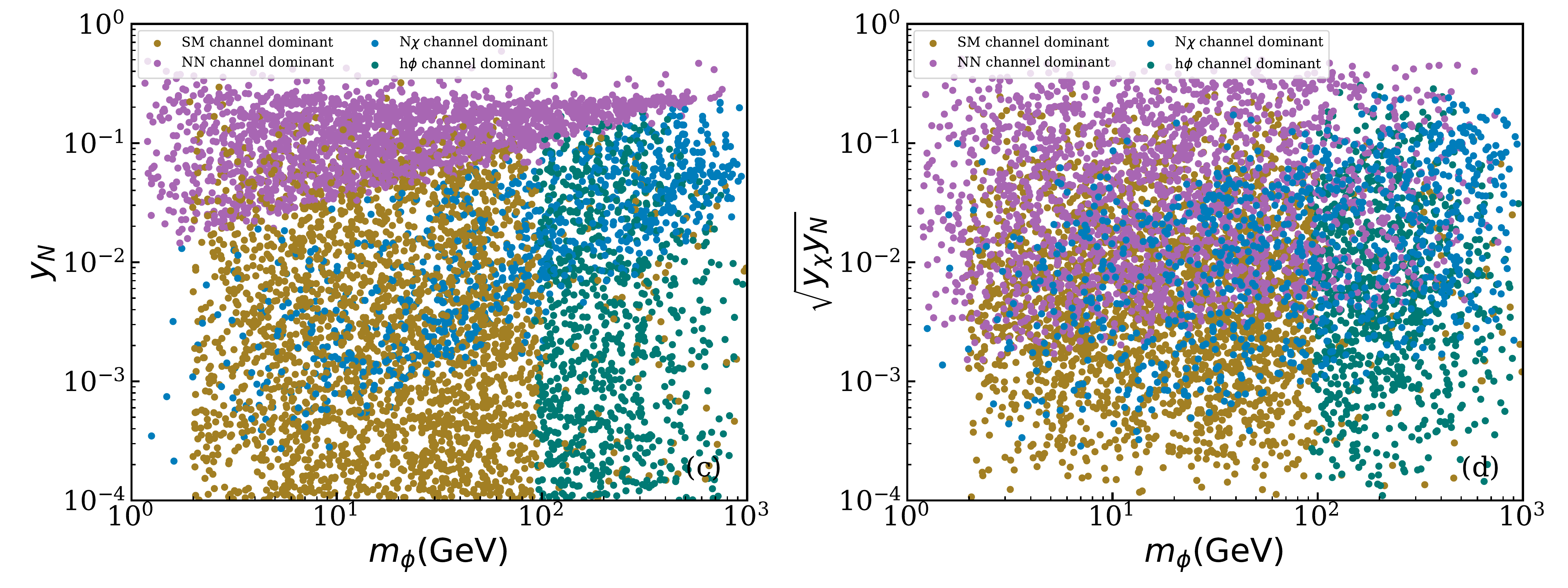}
	\includegraphics[width=1\linewidth]{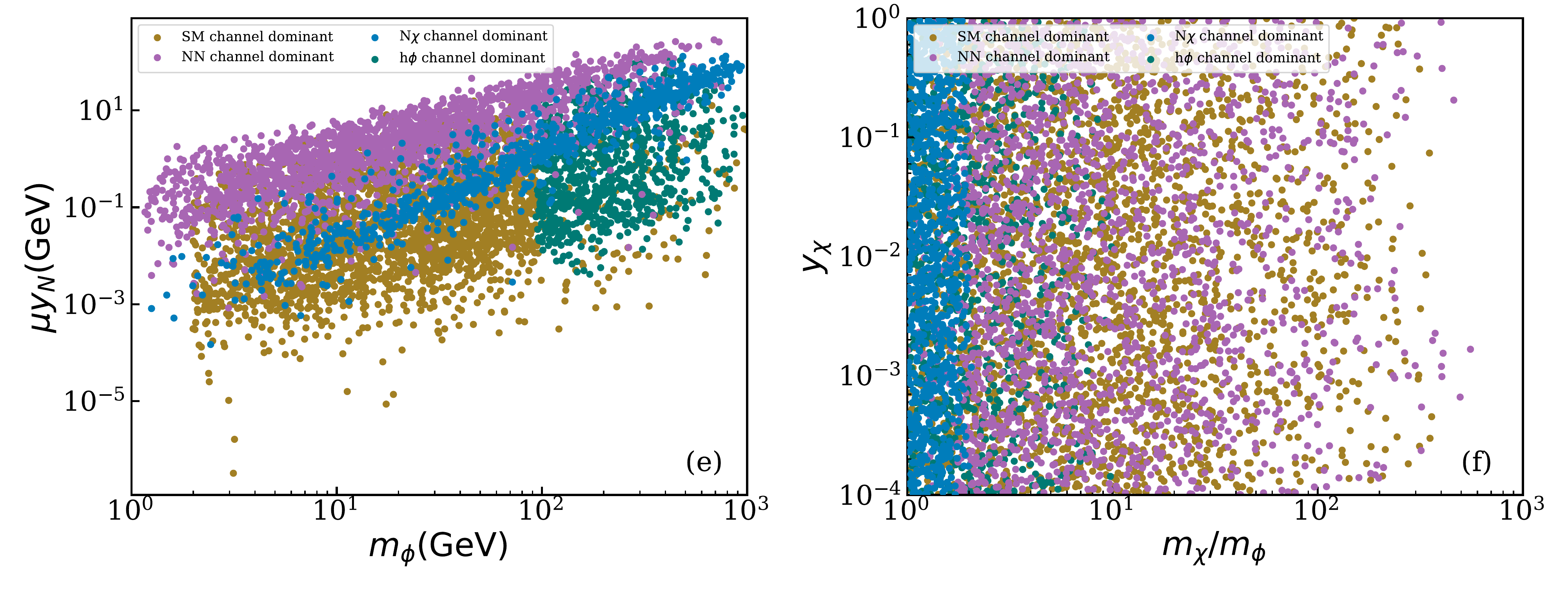}
	\end{center}
	\caption{ Distributions of samples with correct relic density for scalar dark matter. The purple, blue, yellow and green points denote that the dominant annihilation channel is $\phi\phi\to NN$, $\phi\phi\to N\chi$, $\phi\phi\to$ SM, and $\phi\phi\to h \phi$ respectively.}
	\label{Rs}
\end{figure}

Samples with correct relic density in the $3\sigma$ range of the Planck value are kept for later study. Survived samples are then classified by the dominant annihilation channels. The results are shown in Figure~\ref{Rf} and Figure~\ref{Rs} for the case of fermion and scalar dark matter respectively. When $\chi\chi\to NN$ is the dominant channel, it is clear that a lower bound on $y_N$ exists, which is approximately $y_N\gtrsim (m_\chi/10^4~\text{GeV})^{1/2}$. Including the contribution of semi-annihilation channel $\chi\chi\to N\chi$ would allow $y_N$ to be about two orders of magnitude smaller. Similarly, the lower limit $y_\chi\gtrsim m_\chi/10^4~\text{GeV}$ should be satisfied when $\chi\chi\to N\chi$ is the dominant channel. It is also clear that for large enough $y_\chi$, i.e., $y_\chi\gtrsim0.3$, the semi-annihilation channel will always have the largest contribution. These two channels are well separated in the $y_N-y_\chi$ plane, namely $\chi\chi\to N\chi$ is the leading one when $y_\chi\gtrsim0.6y_N$.  As shown in Figure \ref{ff} (b), there is a $s$-channel for the $\chi\chi\to N\chi$ annihilation. Therefore, the contribution of this channel is enhanced when $m_\phi\simeq2m_\chi$, which leads to the deep cusp in panel (d) of Figure \ref{Rf}. Apart from the resonance region, the larger the ratio $m_\phi/m_\chi$ is, the bigger the factor $\sqrt{y_\chi y_N}$ is required.

For the scalar dark matter, $\lambda_{H \phi}\lesssim0.1$ is enough to realize the correct relic density when 10 GeV $\lesssim m_\phi\lesssim10^3$ GeV. The sharp dip around $m_\phi\sim m_h/2$ corresponds to the on-shell production of $h$ in the $s$-channel, where $\lambda_{H \phi}$ can be as small as $10^{-4}$. The SM channel dominant samples via the Higgs portal distribute mainly on the upper edge of the allowed region of $\lambda_{H \phi}$. For the scalar semi-annihilation dominant samples, $\lambda_{H \phi}\gtrsim10^{-3}$ is required. The lower limit on $\lambda_{H \phi}$ for the $h\phi$ dominant channel grows as $m_{\phi}$ increases. So the contribution of the $\phi\phi\to h \phi$ channel is only important in the range of $[10^2,10^3]$  GeV. Since the $h\phi$ channel also involves the cubic coupling $\mu$, the $h\phi$ dominant samples have the largest value of $\mu\lambda_{H \phi}$ in the allowed region. For the secluded channel $\phi\phi\to NN$ and the semi-annihilation channel $\phi\phi\to N\chi$, lower limits on $y_N$ are required, which is similar to the fermion dark matter. However, due to additional contributions from the $\phi\phi\to \text{SM}$ and $\phi\phi\to h\phi$ channels, $y_N$ can be as small as $10^{-4}$ when $m_\phi$ is above $\mathcal{O}(10)$ GeV, which is different from the fermion dark matter. In panels (d) and (e) of Figure \ref{Rs}, we show the corresponding parameters that are involved in the $\phi\phi\to N\chi$ channel. Approximate lower limits on the factors $\sqrt{y_\chi y_N}$ and $\mu y_N$ exist for  $\phi\phi\to N\chi$ dominant samples. To make sure this semi-annihilation channel is kinematically allowed, $m_\chi/m_\phi\lesssim(2m_\phi-m_N)/m_\phi\lesssim2$ should be satisfied, which is depicted in panel (f). There is also a fake upper limit $m_\chi/m_\phi<10$ for the $\phi\phi\to h \phi$ dominant samples, because it happens to be $m_\chi^\text{max}\simeq10^3$~GeV and $m_\phi^\text{min}\simeq10^2$ GeV.

\section{Higgs Invisible Decay}\label{SEC:HID}

In principle, the light sterile neutrino could induce additional Higgs decay mode $h\to \nu N$ when $m_N\lesssim m_h$. However, with the seesaw predicted Yukawa coupling $y_\nu\sim\sqrt{2 m_\nu m_N}/v\sim\mathcal{O}(10^{-6})$, the corresponding decay width is heavily suppressed. Meanwhile, for a light enough dark matter, it can contribute to the Higgs invisible decay. The corresponding branching ratio has been constrained by the ATLAS experiment with \cite{ATLAS:2020kdi}:
\begin{align}
	\mathrm{Br}_{\mathrm{inv}} & = \frac{\Gamma_{\mathrm{inv}}}{\Gamma_{\mathrm{inv}}+\Gamma_{\mathrm{SM}}}<0.11,
\end{align}
where $\Gamma_{\mathrm{SM}}\simeq4$ MeV is the standard Higgs width.
The theoretical Higgs invisible decay widths into the dark matter are given by \cite{Escudero:2016ksa}:
\begin{eqnarray}
	\Gamma(h \rightarrow \phi \phi) & = & \frac{\lambda_{H \phi}^{2} v^{2}}{8 \pi m_{h}} \sqrt{1-\frac{4 m_{\phi}^{2}}{m_{h}^{2}}},
  \\
	\Gamma(h \rightarrow \bar{\chi} \chi)&=&\frac{m_h(\lambda_{H \chi}^{e f f})^2}{8 \pi}\left(1-\frac{4 m_{\chi}^2}{m_h^2}\right)^{3 / 2},
\end{eqnarray}
where the one-loop effective $h\bar{\chi}\chi$ coupling is
\begin{eqnarray}\label{Eq:hxx}
	\lambda_{H \chi}^{e f f} & = & \lambda_{H \phi} \frac{y_{N}^{2}}{16 \pi^{2}} \frac{m_{N}}{(m_{\phi}^{2}-m_{N}^{2})^{2}}\left(m_{\phi}^{2}-m_{N}^{2}+m_{N}^{2} \log \frac{m_{N}^{2}}{m_{\phi}^{2}}\right).
\end{eqnarray}

\begin{figure}
	\begin{center}
		\includegraphics[width=16cm,height=6cm]{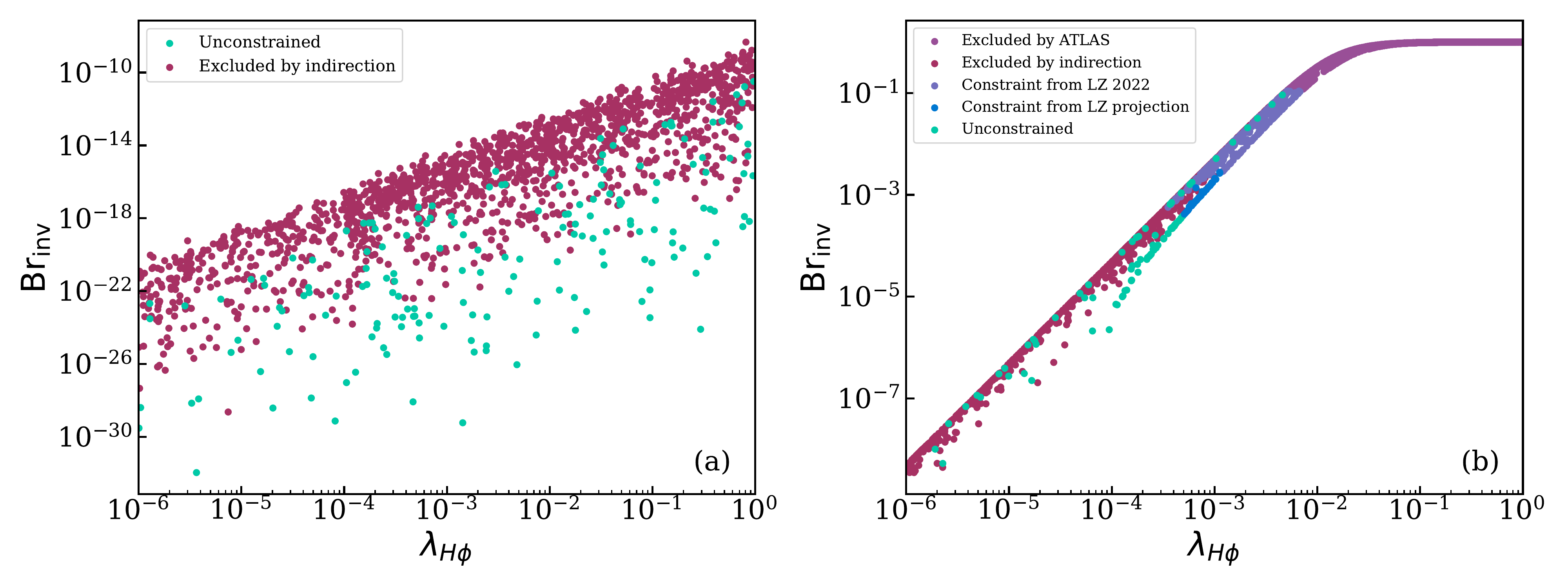}
		\includegraphics[width=16cm,height=6cm]{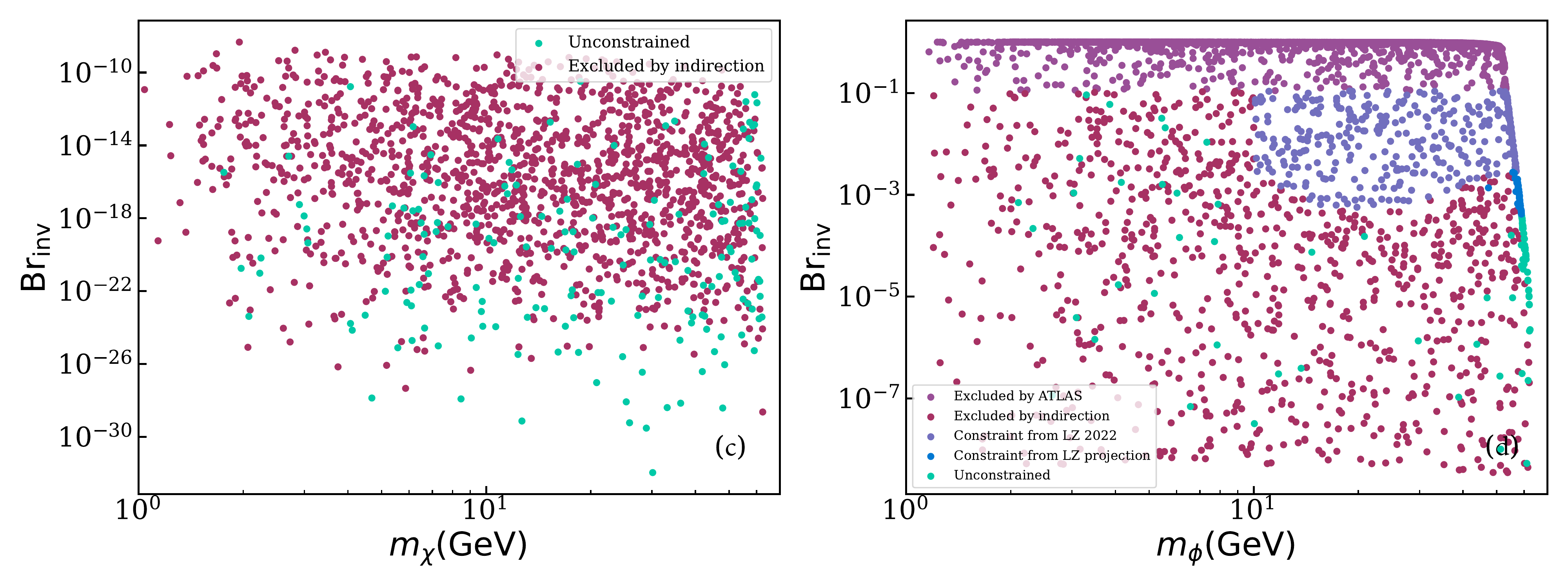}
		\includegraphics[width=16cm,height=6cm]{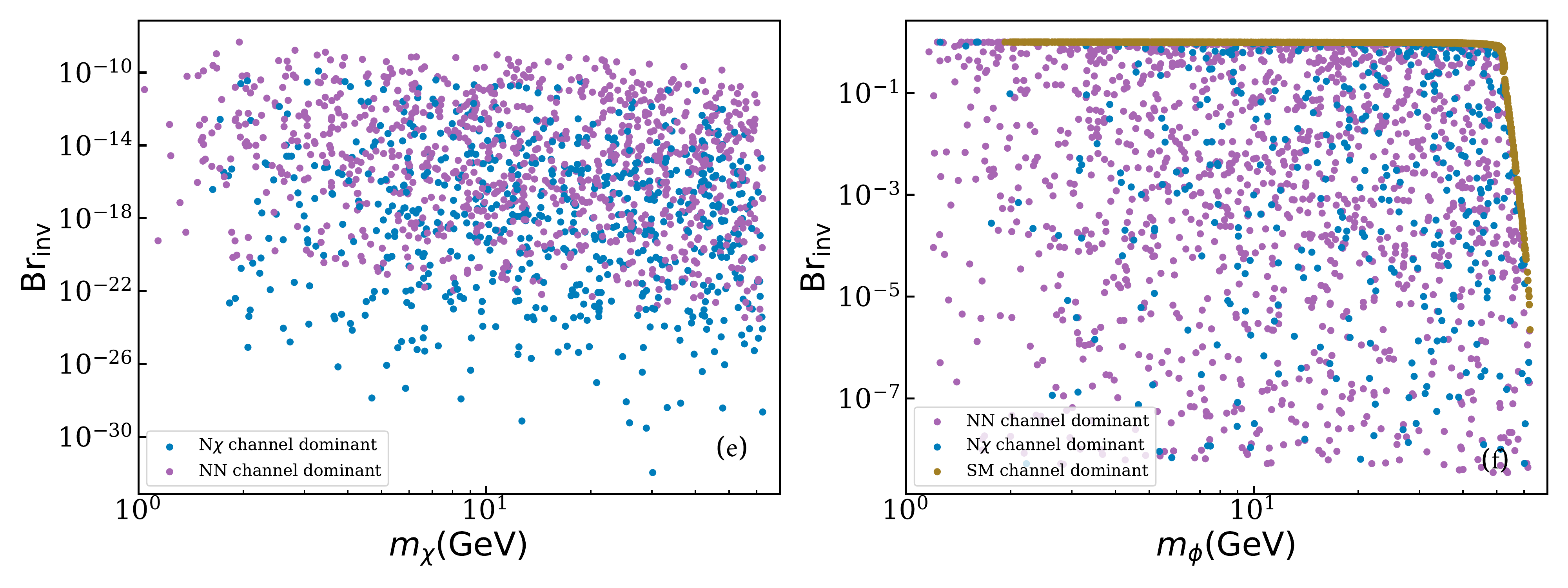}
	\end{center}
	\caption{Branching ratios of invisible Higgs decay induced by fermion (left panels) and scalar dark matter (right panels). In panels (a)-(d), the purple, slate blue, and red points are excluded by ATLAS search of Higgs invisible decay \cite{ATLAS:2020kdi}, direct detection of LZ \cite{LZ:2022ufs}, and indirect detection \cite{Campos:2017odj}, respectively. The blue points satisfy all current constraints but are within the reach of the future LZ experiment \cite{LZ:2015kxe}, meanwhile the green points are unconstrained. In panels (e) and (f), we also classify the samples according to the dominant annihilation channels. Labels of the samples are the same as in Figure \ref{Rf} and Figure \ref{Rs}.}
	\label{FIG:fsmbr}
\end{figure}

Figure~\ref{FIG:fsmbr} shows the theoretical branching ratios of Higgs invisible decay induced by dark matter and various constraints. For fermion dark matter, the branching ratio of Higgs invisible decay is below $10^{-10}$, which is because of the loop suppression of effective Higgs coupling $\lambda_{H \chi}^{eff}$.  The predicted branching ratio is proportional to the coupling $\lambda_{H \phi}$. For maximum Br$_\text{inv}\simeq10^{-10}$, the dominant annihilation channel is found to be $NN$. Meanwhile, for extremely tiny Br$_\text{inv}\lesssim10^{-22}$, the $N\chi$ channel is the dominant one. Because the future HL-LHC could only probe Br$_\text{inv}\gtrsim0.033$ \cite{ATLAS:2018jlh},  this negligible tiny branching ratio induced by fermion dark matter is thus far beyond the reach of direct collider measurement. As will show later, most samples in the light dark matter region below about 60 GeV are excluded by indirect search.

The scenario for scalar dark matter is quite different, where Br$_\text{inv}$ could be the dominant decay channel of Higgs when $\lambda_{H \phi}>0.1$. Under current constraints, the region with $m_\phi\lesssim53$ GeV and Br$_\text{inv}>0.11$ is excluded by the ATLAS experiment, which corresponds to $\lambda_{H \phi}\gtrsim10^{-2}$ disallowed in this region. According to panel (f), most excluded samples in this region are dominantly annihilation via the $\phi\phi\to$ SM channel. For $m_\phi\gtrsim10$ GeV, the direct detection experiment as LZ \cite{LZ:2022ufs} sets a more stringent constrain than the Higgs invisible decay, where Br$_\text{inv}\gtrsim10^{-3}$ and $\lambda_{H \phi}\gtrsim4\times10^{-4}$ might be excluded. Such a small branching ratio is also beyond the reach of future HL-LHC \cite{ATLAS:2018jlh}. As for scalar dark matter lighter than 10 GeV, the Higgs invisible decay leads to more strict constrain than direct detection. The annihilation channels of samples in the region which escapes the limits from Higgs invisible decay and direct detection  are $\phi\phi\to NN$ and $\phi\phi\to N\chi$. Although this region is also tightly constrained by indirect detection,  there are still some samples that satisfy all current constraints. We have checked that most of these light-mass survived samples annihilate via $\phi\phi\to N\chi$ with the special requirement $2m_\phi\lesssim m_N+m_\chi$. So if HL-LHC discovers a relatively large Br$_\text{inv}$, the dark matter candidate should be a scalar with mass around a few GeV, meanwhile the sterile neutrino is also at the GeV scale.

\section{Indirect Detection}\label{SEC:ID}

\begin{figure}
	\begin{center}
		\includegraphics[width=16cm,height=6cm]{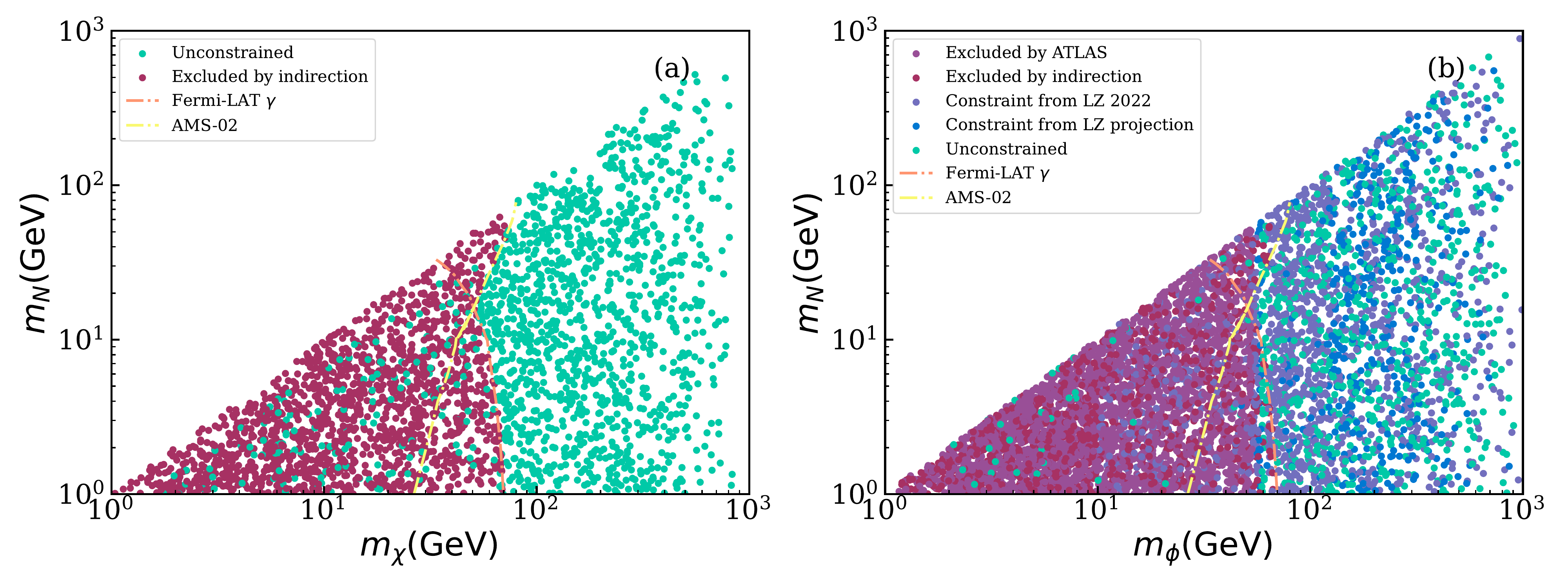}
		\includegraphics[width=16cm,height=6cm]{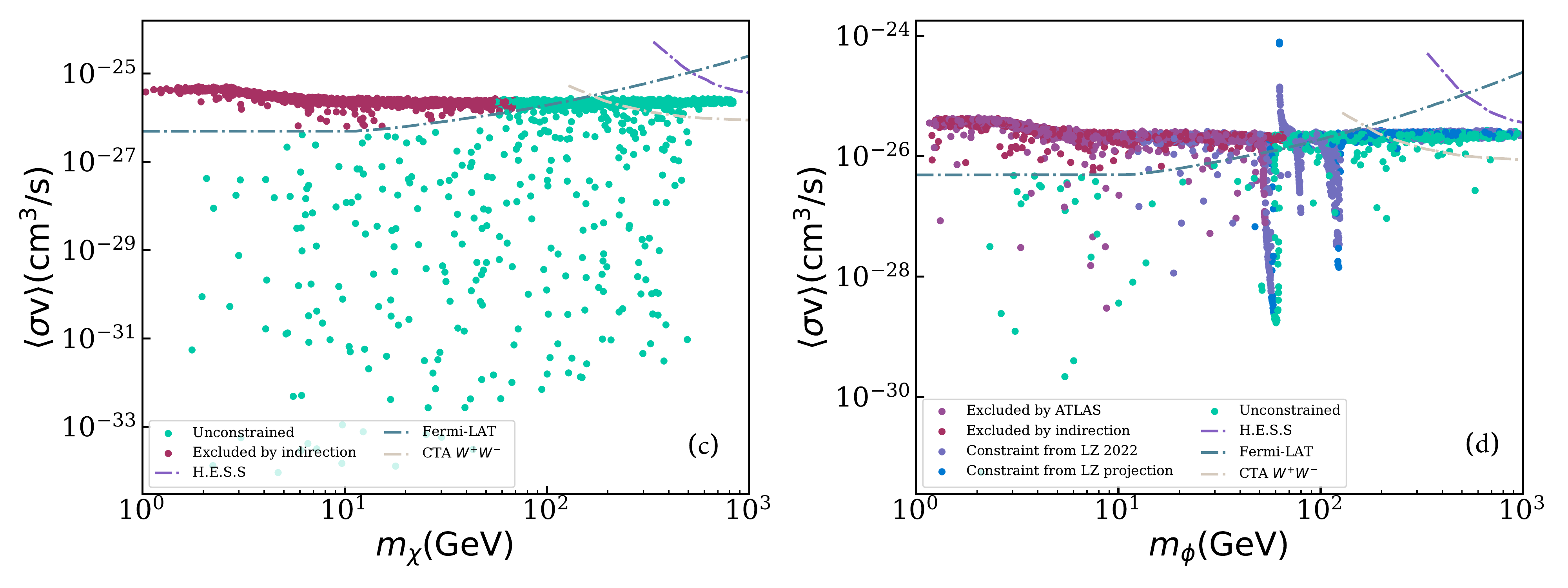}
		\includegraphics[width=16cm,height=6cm]{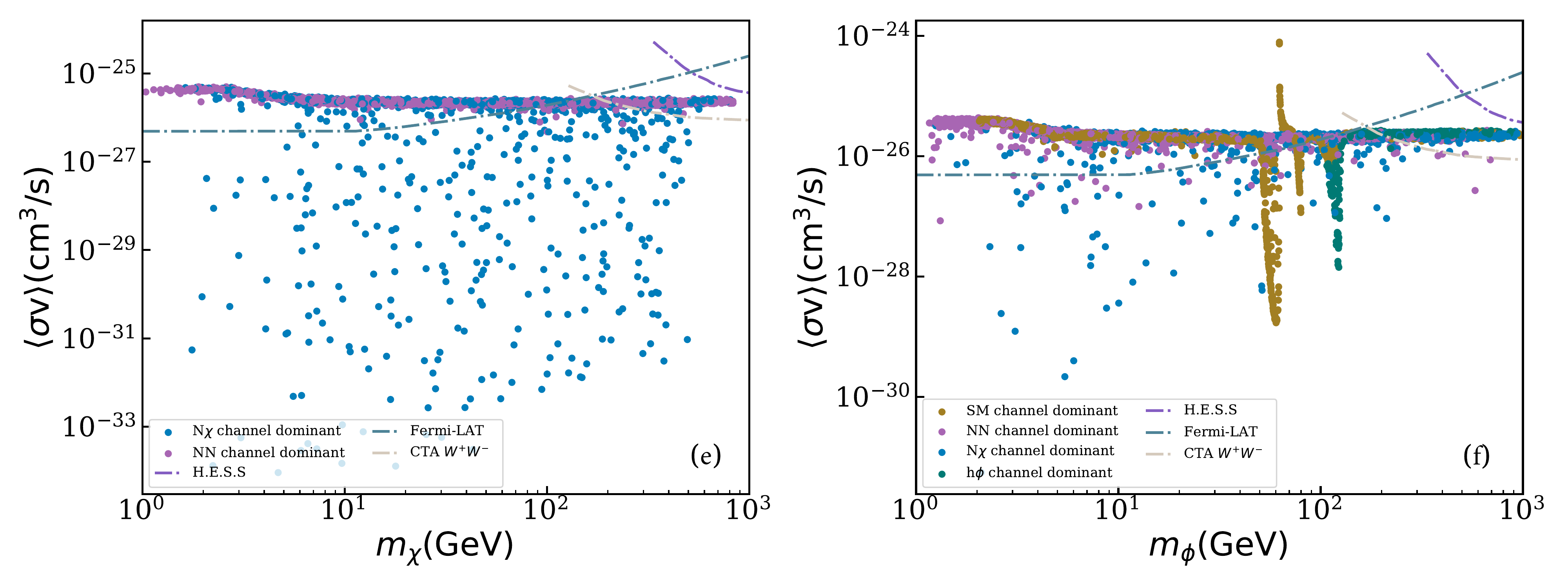}
	\end{center}
	\caption{Exclusion limits of indirect detection experiments. In panels (a) and (b), the yellow and orange curves represent the bounds of the antiproton-to-proton flux ratio from AMS-02 \cite{AMS:2016oqu} and gamma-rays in the Milky Way dSphs from FermiLAT \cite{Fermi-LAT:2015att} with typical thermal annihilation cross section $\langle \sigma v \rangle=2.2\times10^{-26}~\text{cm}^3\text{s}^{-1}$. In panels (c) to (f), the blue and purple curves illustrate the limits on annihilation cross section from Fermi-LAT and H.E.S.S. \cite{Campos:2017odj}. The yellow line shows the sensitivity of future CTA experiment for the $W^+W^-$ annihilation modes \cite{CTA:2015yxo}. Other labels are the same as in Figure \ref{FIG:fsmbr}.}
	\label{sigmav}
\end{figure}

The indirect detection experiments aim to search for various types of particles produced in dark matter annihilation. The differential flux arising from the annihilation of dark matter is calculated as 
\begin{equation}
	\frac{d\Phi}{dE}= \frac{1}{4\pi} \frac{\langle \sigma v \rangle}{2 m_{\mathrm{DM}}^2} \frac{dN}{dE}\cdot\int_{\Delta \Omega} d\Omega \int \rho_{\mathrm{DM}}^2(s)ds,
\end{equation}
where $\rho_{\mathrm{DM}}$ is the dark matter density of the observed object. The energy spectrum $dN/dE$ describes the distribution of observed particles from dark annihilation. According to previous studies, current indirect limits could constrain dark matter mass below about 50 GeV \cite{Batell:2017rol}. In this region, the dominant annihilation channels for both fermion and scalar dark matter are $NN$ and $N\chi$ final states, after imposing the constraint from Higgs invisible decay. The resulting spectrum $dN/dE$ depends on both the masses $m_{\text{DM}},m_N$ and decay modes of sterile neutrino $N$.  With light $m_N<m_W$, the three-body decay width via off-shell $W$ and $Z$ can be estimated as
\begin{equation}
	\Gamma_N\approx \frac{G_f^2}{192\pi^3} |\theta_\alpha|^2 m_N^5,
\end{equation}
where $\theta_\alpha (\alpha=e,\mu,\tau)$ describes the mixing angle between sterile and active neutrinos for different flavors. For a heavier sterile neutrino, the two-body decays $N\to W^\pm \ell^\mp,Z\nu, h\nu$ are the dominant channels. The continuum spectra of muon flavor $N$ is similar to the electron flavor scenario, while the tau flavor $N$ produces a slightly stronger gamma-ray spectrum \cite{Campos:2017odj}. In this paper, we consider an electron flavor $N$ for a conservative study.

Using the observed spectrum, the exclusion limits on dark matter annihilation can be derived by performing a likelihood analysis, although the large astrophysical uncertainties would affect these limits. In Figure \ref{sigmav}, we show the indirect detection constraints from the antiproton observations of AMS-02 and the gamma-rays observations of Fermi-LAT \cite{Batell:2017rol}. For light $m_N\sim\mathcal{O}$(GeV), the Fermi-LAT experiment could exclude  $m_\text{DM}\lesssim60$ GeV. Meanwhile for heavier $m_N$, the AMS-02 result would exclude $m_\text{DM}\lesssim80$ GeV. The combination of these two limits excludes most samples in the region below $m_\text{DM}\lesssim50 \mathrm{GeV}$. It is notable that these two limits in Figure \ref{sigmav} (a) and (b) are obtained with fixed thermal annihilation cross section $\langle \sigma v \rangle=2.2\times10^{-26}~\text{cm}^3\text{s}^{-1}$. In panel (c)-(f), we show the theoretically predicted annihilation cross section for indirect detection,  where the cross sections of the semi-annihilation process as $\chi\chi/\phi\phi\to N\chi$, $\phi\phi\to h\phi$ are multiplied by a factor of $1/2$. Since the annihilation cross section today can be much smaller than the thermal target $\langle \sigma v \rangle=2.2\times10^{-26}~\text{cm}^3\text{s}^{-1}$, we require that the samples are excluded by indirect detection when the corresponding cross sections are also above the Fermi-LAT limit on $\langle \sigma v \rangle$.

In the low mass region below 50 GeV, the two annihilation channels of fermion dark matter lead to quite different results. For the $\chi\chi\to NN$ dominant samples, the corresponding annihilation cross sections are at the typical value $\langle \sigma v \rangle=2.2\times10^{-26}~\text{cm}^3\text{s}^{-1}$, so these samples are excluded by indirect detection. However for the $\chi\chi\to N\chi$ dominant samples, due to the existence of $s$-channel contribution via the dark scalar $\phi$, the annihilation cross section could be much smaller than $2.2\times10^{-26}~\text{cm}^3\text{s}^{-1}$ when $2m_\chi\simeq m_\phi$. In this special scenario, the $\chi\chi\to N\chi$ dominant samples also satisfy the indirect constraints. For the scalar dark matter, although there are three annihilation modes in these low mass region, the $\phi\phi\to\text{SM}$ channel is tightly constrained by Higgs invisible decay and direct detection. 

\begin{figure}
	\begin{center}
		\includegraphics[width=16cm,height=6cm]{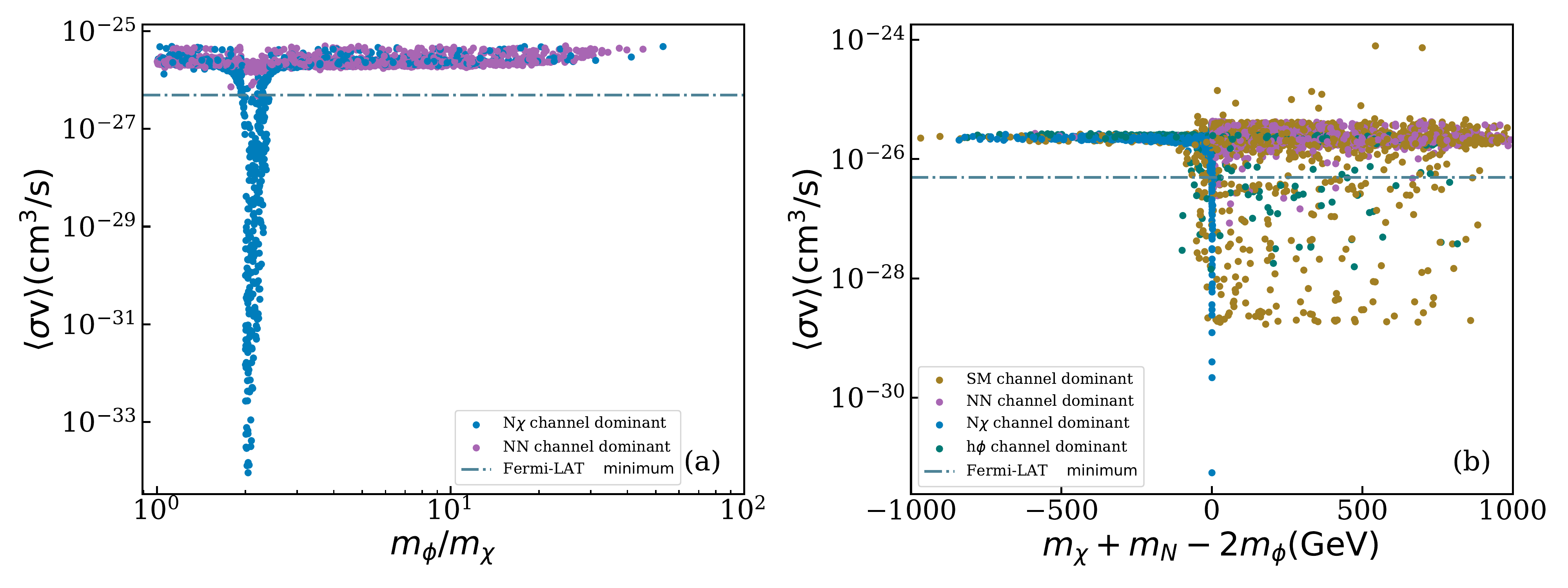}
	\end{center}
	\caption{Annihilation cross section for indirect detection. The horizontal dash-dotted line is the minimum value of the Fermi-LAT limit in Figure \ref{sigmav}}
	\label{sv}
\end{figure}

Similar to the fermion dark matter, most of the $\phi\phi\to NN$ dominant samples could be excluded by indirect detection in the low mass region, while some of the $\phi\phi \to N\chi$ dominant samples are still allowed. Although there is no on-shell $s$-channel contribution of the $\phi\phi \to N\chi$ channel, we find that the allowed samples satisfy $2m_\phi\lesssim m_\chi+m_N$ as shown in Figure \ref{sv}. So these samples fall into the forbidden region, where the non-relativistic velocity of dark matter today can not overcome the mass splitting $m_\chi+m_N-2m_\phi$ \cite{DAgnolo:2015ujb}. The scalar dark matter annihilates into SM particles also has on-shell $s$-channel contributions when $2m_\phi\simeq m_h$.  Meanwhile, the scalar semi-annihilation $\phi\phi\to h\phi$ meets the forbidden condition when $m_\phi\lesssim m_h$. The resulting annihilation cross sections of these two kinds are much smaller than the typical value $\langle \sigma v \rangle=2.2\times10^{-26}~\text{cm}^3\text{s}^{-1}$, so these samples are hard to probe by indirect detection. 

For dark matter around the TeV scale, the most stringent constraint is from H.E.S.S. observation \cite{HESS:2011zpk}. Currently, no samples could be excluded by H.E.S.S.. If there is no positive signal at future direct detection experiments, the scalar semi-annihilation $\phi\phi\to h\phi$ would be excluded. So in this heavy mass region, we only consider the sterile neutrino portal annihilation channels $\phi\phi/\chi\chi\to NN$. Because the photon spectrum from $NN$ final state is similar to $W^+W^-$ final state, we also show the future limits from the Cherenkov Telescope Array (CTA) \cite{CTA:2015yxo}, which will cover most samples above 200 GeV.

\section{Direct Detection}\label{SEC:DD}

In this model, the dark matter scatters off the atomic nucleus elastically via the $t$-channel exchange of Higgs boson $h$. For scalar dark matter, this scattering happens at tree-level, while for fermion dark matter, it is induced at one-loop level \cite{Escudero:2016ksa}. The dark matter direct detection experiments measure the nuclear recoil energy and set constraints on the dark matter-nucleon scattering cross section. Until now, no concrete signal is observed by direct detection experiments, such as PandaX-4T \cite{PandaX-4T:2021bab}, XENONnT \cite{XENON:2023sxq}, and LZ \cite{LZ:2022ufs}. In this paper, we consider the most stringent limit from LZ and Darkside-50 \cite{DarkSide-50:2022qzh} at present and the future projected limit from LZ \cite{LZ:2015kxe}. For light dark matter below about 10 GeV, the Darkside-50 experiment sets the most stringent limit, which excludes $\sigma_{\rm SI}\gtrsim10^{-43}~\text{cm}^2$. For heavier dark matter, the LZ limit is the most tight one, where the minimum is at $m_\text{DM}=30$ GeV with $\sigma_{\rm SI}=5.9\times10^{-48}~\text{cm}^2$.

Within the context of the Higgs-portal effective scenarios \cite{Arcadi:2021mag}, the spin-independent cross section for dark matter collision with nucleon can be expressed as 
\begin{eqnarray}
	&& \sigma_{\rm SI}^{\phi n} = \frac{\lambda_{H\phi}^2}{  \pi m_h^4} \frac{m_n^4  f_n^2}{ (m_{\phi} + m_n)^2} \;,  \\
	&& \sigma_{\rm SI}^{\chi n} = \frac{(\lambda_{H\chi}^{eff})^2}{\pi m_h^4} \frac{m_n^4 m_\chi^2  f_n^2}{ (m_\chi + m_n)^2} \;.
\end{eqnarray}
The nucleon mass is denoted as $m_n$, and the parameter $f_n \simeq 0.3$ are used to parameterize the Higgs-nucleon interactions \cite{Cline:2013gha}. The effective coupling $\lambda_{H\chi}^{eff}$ is calculated in Equation \eqref{Eq:hxx}.

\begin{figure}
	\begin{center}
		\includegraphics[width=1\linewidth]{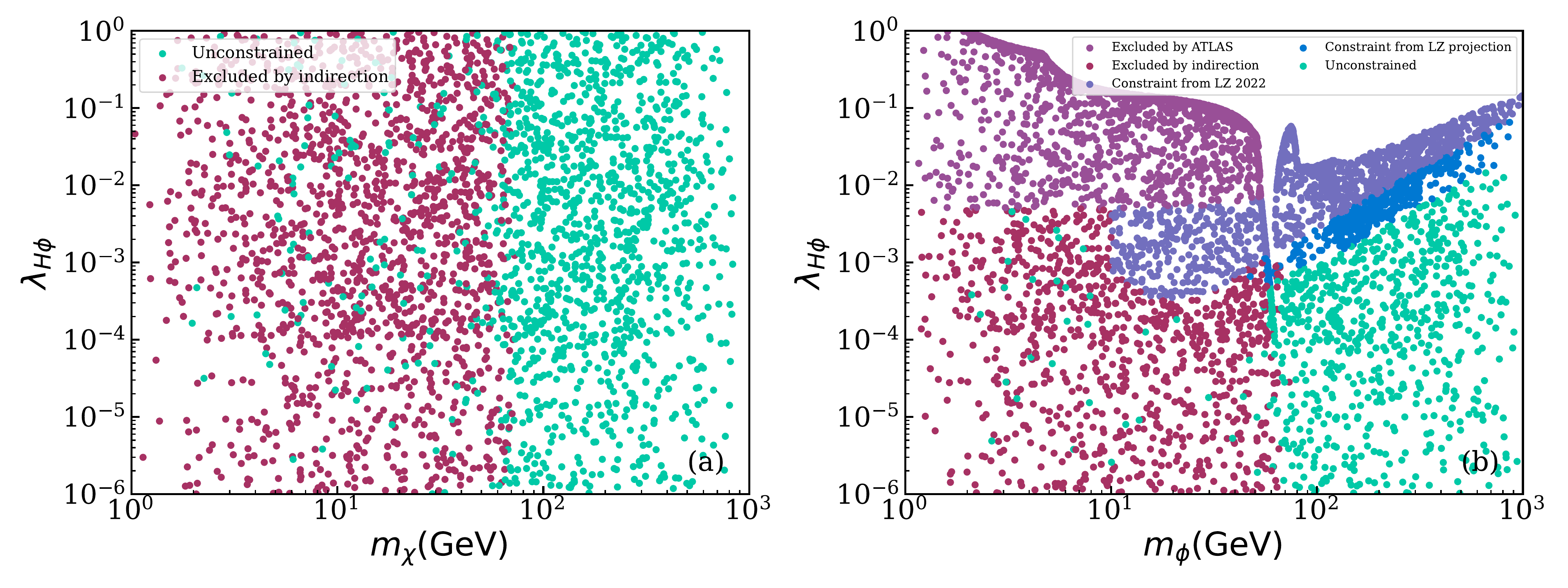}
		\includegraphics[width=1\linewidth]{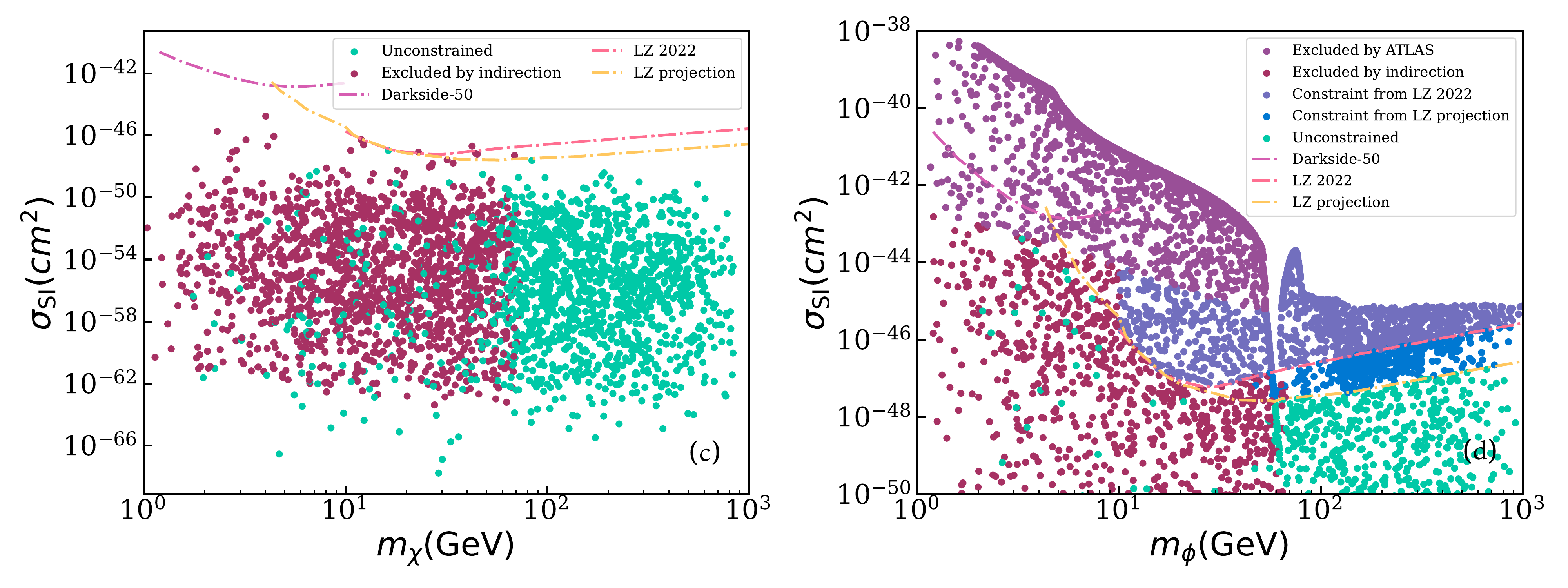}
		\includegraphics[width=1\linewidth]{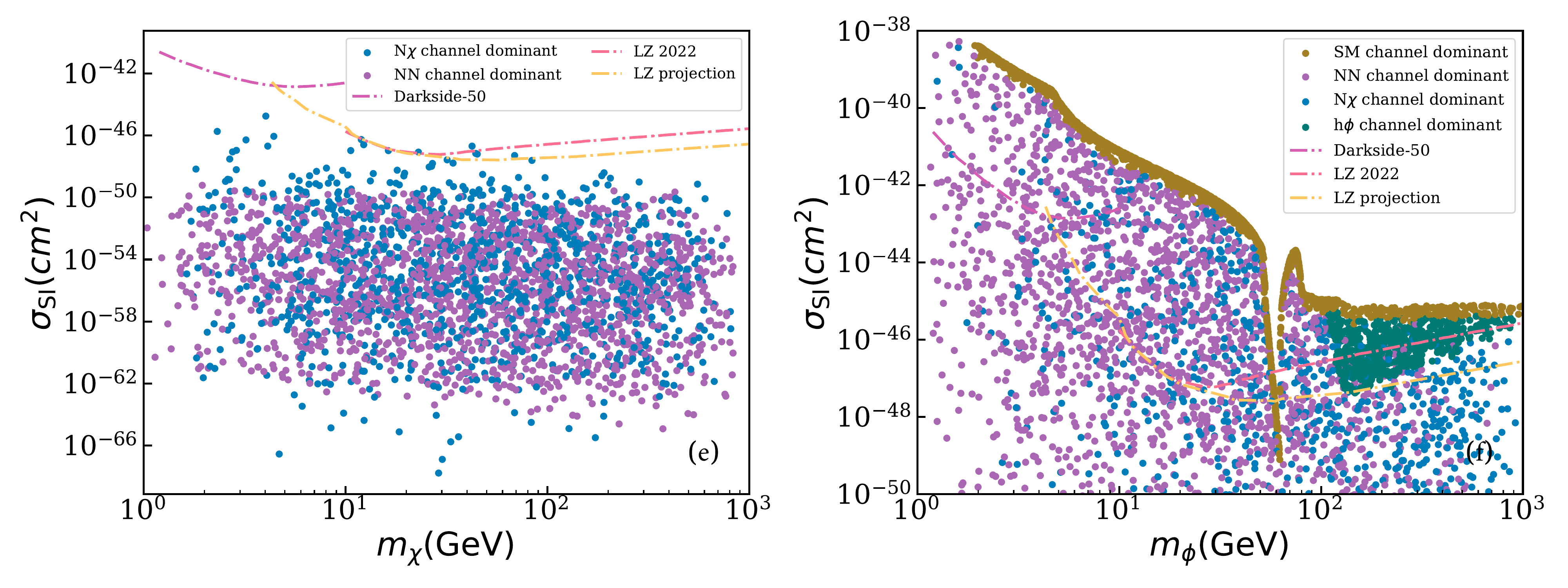}
	\end{center}
	\caption{ The predicted spin-independent cross section and various exclusion limits. The purple and red lines are the current exclusion limits from Darkside-50 and LZ respectively. The yellow line is the projected LZ limit. Other labels are the same as in Figure \ref{FIG:fsmbr}.}
	\label{fsms}
\end{figure}

The scanned results are shown in Figure \ref{fsms} for both fermion and scalar dark matter.  Because the scattering cross section for fermion dark matter is suppressed by the one-loop factor, the predicted values are usually far below current experimental limits. Within the parameter space scanned in Equation~\eqref{Eq:PR}, even the future projection of the LZ experiment could not have a positive signature when taking into account the limits from indirect detection. We also find that samples with relatively large scattering cross section $\sigma_{\rm SI}\gtrsim10^{-50}~\text{cm}^2$ are dominant by the semi-annihilation channel  $\chi\chi\to N\chi$.  In principle, increasing the maximum values of related coupling $\lambda_{H\phi}$ and $y_N$ to the perturbation limits could lead to the predicted scattering cross section above current limits \cite{Escudero:2016ksa}. These detectable samples are expected to annihilate via  $\chi\chi\to N\chi$ channel in this model.

As for the scalar dark matter, the coupling $\lambda_{H \phi}$ induces quite a large scattering cross section with correct relic density. The direct detection experiments, such as LZ, could now exclude samples with $\lambda_{H \phi}\gtrsim4\times10^{-4}$. As already discussed in Section \ref{SEC:HID}, the Higgs invisible decay has excluded $\lambda_{H \phi}\gtrsim10^{-2}$ for $m_\phi<m_h/2$. It is clear in panel (d) of Figure \ref{fsms} that the exclusion limit from Higgs invisible decay is more stringent than the direct limit from Darkside-50, therefore we do not consider the Darkside-50 limit in this study. For $m_\phi\gtrsim10$~GeV, the direct limit from LZ is over two orders of magnitudes tighter than the Higgs invisible limit. Since the indirect detection has already excluded most samples with $m_\phi\lesssim50$ GeV, the future projection of LZ is also hard to probe the allowed samples in this region. Together with panel (f) of Figure~\ref{fsms}, the only viable region of $\phi\phi\to \text{SM}$ dominant samples is the narrow resonance region at $m_\phi\lesssim m_h/2$. The predicted cross section of SM dominant samples can be as small as $\sigma_{\rm SI}\sim10^{-49}~\text{cm}^2$, which is also beyond the reach of future LZ.

For heavier scalar dark matter above 100 GeV, only the direct detection experiments set the corresponding limit now.  The $\phi\phi\to \text{SM}$ dominant samples lead to the largest predicted cross section $\sigma_{\rm SI}\simeq10^{-45}~\text{cm}^2$, which is disfavored by the current LZ  limit when $m_\phi<1$ TeV. It is still possible for $\phi\phi\to \text{SM}$ dominant samples with $m_\phi$ above 1 TeV \cite{Athron:2018ipf}, but is out of the parameter space we scanned in Equation \eqref{Eq:PR}. The new contribution of the semi-annihilation $\phi\phi\to h\phi$ channel would induce a smaller scattering cross section. Since this semi-annihilation channel also involves the cubic coupling, the stability and unitarity bounds $\mu<3 m_\phi$ then lead to a lower bound for the predicted cross section. For instance, the minimum predicted cross section is about $3\times10^{-47}~\text{cm}^2$ with $m_\phi\sim130$ GeV. Once $\phi\phi\to h\phi$ is kinematically allowed, this lower bound on $\sigma_{\rm SI}$ increases as $m_\phi$ is larger. Under the current LZ limit, some of the $\phi\phi\to h\phi$ dominant samples are still allowed. In the future, the projected LZ limit could probe all the $\phi\phi\to h\phi$ dominant samples. So if there is still no positive signature at future LZ, the allowed samples will be dominant by $\phi\phi\to NN$ and $\phi\phi\to N\chi$ channels. We then expect observable signatures at indirect detection experiments from these two channels for most of the allowed parameter space. 
 
\section{Conclusion}\label{SEC:CL}

Besides generating tiny neutrino mass via the type-I seesaw mechanism, the electroweak scale sterile neutrino $N$ can also mediate the interaction between the dark sector and the standard model. Beyond the simplest $Z_2$ symmetry, we extend the sterile neutrino portal dark matter with $Z_3$ symmetry  in this paper. We introduce a scalar singlet $\phi$ and a fermion singlet $\chi$ to the dark sector. Under the dark $Z_3$ symmetry, the dark sector transforms as $\chi\to e^{i2\pi/3}\chi, \phi\to e^{i2\pi/3}\phi$, while the standard model particles and the sterile neutrinos transform trivially. In this paper, we consider WIMP dark matter for both scalar and fermion scenarios. The $Z_3$ symmetry introduces two new interactions, i.e., $y_\chi \phi \overline{\chi^{c}} \chi$ and $\mu\phi^3/2$, which lead to semi-annihilation channels as $\chi\chi\to N\chi$ and $\phi\phi\to h\phi$.

For the fermion dark matter $\chi$, the annihilation channels are secluded $\chi\chi\to NN$ and the semi-annihilation $\chi\chi\to N\chi$. Because the effective $h\bar{\chi}\chi$ coupling is induced at the one-loop level, the contribution of fermion dark matter to Higgs invisible decay is negligible tiny. The resulting dark matter-nucleon scattering cross section is also beyond future experimental reach. Currently, the indirect detection could exclude most of the samples with $m_\chi\lesssim50$ GeV. In the future, the CTA experiment is expected to probe the high mass region. However, due to the  $s$-channel contribution of dark scalar $\phi$ to the semi-annihilation $\chi\chi\to N\chi$, the corresponding annihilation cross section is much smaller than the usual thermal value $\langle \sigma v\rangle =2\times10^{-26}~\text{cm}^3\text{s}^{-1}$ when $2m_\chi\simeq m_\phi$. In this special scenario, even the indirect detection can not have a positive signature.

For the scalar dark matter $\phi$, there are four kinds of annihilation channels, i.e., the Higgs portal $\phi\phi\to \text{SM}$, the secluded channel $\phi\phi\to NN$, and the semi-annihilations $\phi\phi\to N\chi,h\phi$. The direct Higgs portal interaction $\lambda_{H \phi} (H^\dag H) (\phi^\dag \phi)$ generates observable signatures from Higgs invisible decay, indirect and direct detection. Under these constraints, the Higgs portal $\phi\phi\to \text{SM}$ dominant samples are only viable at the resonance region $m_\phi\lesssim m_h/2$ for dark scalar below 1~TeV. The semi-annihilation $\phi\phi\to h\phi$ might be the dominant one in the range of $[10^2,10^3]$ GeV. This channel predicts a lower bound on dark matter-nucleon scattering cross section, which can be fully detected by the future LZ experiment. Meanwhile, the secluded $\phi\phi\to NN$ and semi-annihilation $\phi\phi\to N\chi$ channels can easily satisfy current bounds and are promising at indirect detection experiments. Although the light mass region is tightly constrained,  we find that when the forbidden relation $2m_\phi\lesssim m_\chi+m_N$ is satisfied, the semi-annihilation channel $\phi\phi\to N\chi$ also has a suppressed annihilation cross section for indirect detection.

Compared with the $Z_2$ symmetric model, the new interactions introduced in the $Z_3$ symmetric model enlarge the viable parameter space. For instance, light dark matter below about 50 GeV is entirely excluded by indirect detection in the $Z_2$ symmetric model. However, in the $Z_3$ symmetric model, the semi-annihilation $\chi\chi\to N\chi$ and $\phi\phi\to N\chi$ are still allowed under certain circumstances. Therefore, once light dark matter is discovered, the $Z_3$ symmetric model would be preferred.

\section*{Acknowledgments}
This work is supported by the National Natural Science Foundation of China under Grant No. 11805081 and  11635009, Natural Science Foundation of Shandong Province under Grant No. ZR2019QA021 and ZR2022MA056, the Open Project of Guangxi Key Laboratory of Nuclear Physics and Nuclear Technology under Grant No. NLK2021-07.


\end{document}